\documentclass[twocolumn,trackchanges]{aastex62}

\usepackage{nccmath}
\usepackage{multirow}
\usepackage{amsmath, amstext}
\usepackage{amssymb}
\usepackage{amsmath}
\usepackage{mathtools}
\usepackage{color} 
\usepackage{threeparttable}
\usepackage{float}

\received{April 27th, 2020}
\revised{May 5th, 2020}
\accepted{October 23th, 2020}

\submitjournal{ApJ}

\shorttitle{Variability of Mg~II and Fe~II in 3C~454.3}
\shortauthors{Amaya-Almaz\'an et al.}

\begin{document}

\title{Multiwavelength analysis and the difference in the behavior of the spectral features during the 2010 and 2014 flaring periods of the blazar 3C 454.3}

\correspondingauthor{Ra\'ul A. Amaya-Almaz\'an}
\email{rantonioaa@inaoep.mx, amayaalmazanra@gmail.com}

\author[0000-0002-9443-7523]{Ra\'ul A. Amaya-Almaz\'an}
\affil{Instituto Nacional de Astrof\'isica, \'Optica y Electr\'onica, Luis Enrique Erro $\# 1$, 
Tonantzintla, Puebla 72840, M\'exico}

\author[0000-0002-2558-0967]{Vahram Chavushyan}
\affil{Instituto Nacional de Astrof\'isica, \'Optica y Electr\'onica, Luis Enrique Erro $\# 1$,
Tonantzintla, Puebla 72840, M\'exico}

\author[0000-0002-5442-818X]{Victor M. Pati\~no-\'Alvarez}
\affil{Instituto Nacional de Astrof\'isica, \'Optica y Electr\'onica, Luis Enrique Erro $\# 1$, 
Tonantzintla, Puebla 72840, M\'exico}
\affiliation{Max-Planck-Institut f\"ur Radioastronomie, Auf dem H\"ugel 69,
D-53121 Bonn, Germany}

\begin{abstract}

The flat-spectrum radio quasar 3C~454.3 throughout the years has presented very high activity phases (flares) in which the different wavebands increase their flux dramatically.  In this work, we perform multiwavelength analysis from radio to gamma-rays and study the Mg~II~$\lambda 2798$\AA\ emission line and the UV~Fe~II band from 2008-2018. We found that an increase in the 43 GHz flux density of the quasi-stationary component C, coincides with the estimated time at which a superluminal blob ejected from the radio core (which caused the brightest flare of 2010) collides with the quasi-stationary component (at a projected distance of $\sim4.6$ pc from the radio core). The spectral index different behavior in the first ($5000 < \text{JD}-2450000 < 5600$) and second ($6600 < \text{JD}-2450000 < 7900$) flaring periods suggest changes in the physical conditions. The complex nature of the second period can be a result of a superposition of multiple events at different locations. The Mg~II has an anti-correlation with the UV-continuum while Fe~II correlates positively. Except by the time of the brightest flare of 2010, when both have a strong response at high continuum luminosities. Our results suggest that the dominant gamma-ray emission mechanism for the first flaring period is External Compton. For the second flaring period the seed photons emission region is co-spatial with the gamma-ray emission region. However, a SED study using a multizone jet emission model is required to confirm the nature of each significant flare during the second period.

\end{abstract}

\keywords{galaxies: active -- galaxies: jets -- gamma rays: galaxies -- line: formation -- quasars: emission lines -- quasars: individual: 3C 454.3}

\section{Introduction} \label{sec:intro}

Flat-Spectrum Radio Quasars (FSRQ), a sub-class of blazar-type Active Galactic Nuclei (AGN), display key characteristics of blazars and quasars, very high variability and emission lines. They display variability on a wide range of timescales \citep{Fan2018,Gupta2018} as well. In most FSRQ, the continuum from radio to UV is often dominated by non-thermal emission, which is coming from a jet of relativistic plasma. The jet is pointing towards our line of sight \citep{Urry&Padovani1995} causing Doppler boosting \citep{Sher1968}. The Spectral Energy Distribution (SED) of this type of object has two main components.
The low energy component mostly due to the synchrotron emission and the high energy component due to inverse Compton (IC) scattering \citep[e.g.][]{Bottcher2007,Bottcher2013,Romero2017}. The process of emission through IC can be due to two different physical interactions in the central engine; one of these is called Synchrotron Self-Compton \citep[SSC,][]{Bloom1996} in which the seed photons, to be taken to higher energies by the relativistic electrons, come from the synchrotron emission produced in the jet itself. On the other hand in External inverse Compton \citep[EC,][]{Sikora1994}, like its name alludes, the seed photons come to the jet from external sources like the accretion disk, the broad-line region (BLR), or the dusty torus.

The object of interest 3C~454.3, $z=0.859$, is a blazar and has been classified as an FSRQ due to its high optical variability \citep{Angione1968} and prominent broad emission lines. This source shows structural and flux variability on its parsec scale jet \citep[e.g.][]{Pauliny-Toth1987,Kemball1996,Jorstad2013,Jorstad2017} and correlated variability between different wavebands \citep[e.g.][]{Tornikoski1994,Zhang2018,Zhang2020}. The very high variability of this blazar is a reason for its constant study and observational campaigns in which 3C~454.3 has been monitored throughout all these years.

The location of the gamma-ray emission region in this object has not been completely determined. Several studies have discussed that the location might be close to the black hole inside the BLR \citep{Tavecchio2010,PoutanenStern2010,Hu2015}, while others, that this emission region should be located outside the BLR downstream the jet \citep{Sikora2008,Jorstad2010,Vittorini2014,Coogan2016,Britto2016}. Another important issue that has not reached a consensus is the dominant gamma-ray emission mechanism of 3C~454.3. There are times when the dominant gamma-ray emission mechanism is determined to be EC \citep{Hu2015,Vittorini2014}, however, other studies show indications that SSC is likely to be the dominant mechanism \citep{Kushwaha2016,Rajput2019}. These studies show that 3C~454.3 could be an example of a source with multiple gamma-ray emission regions, as well as, the dominant gamma-ray emission mechanism being able to change.

The link between optical continuum, emission-line variability, and jet kinematics on sub-parsec scales was proposed to be explained by the existence of jet-excited BLR outflowing downstream the jet \citep{Arshakian2010,LeonTavares2010}. \cite{Leon-Tavares2013} presented direct observational evidence of the BLR close to the radio core of the jet and was confirmed in consequent studies \citep{Isler2013,Jorstad2013}. \cite{Arshakian2012} and \cite{Leon-Tavares2015} present schematics in which this additional BLR (BLR material surrounding the jet) is depicted in contrast to the canonical BLR \citep[virialized clouds located in the inner parsec region of the central engine, e.g.][]{Kaspi2005}. The origin of this additional BLR is still under debate. The most likely origin of this clouds is wind from the accretion disk accelerated by the magnetic field of the jet \citep{Perez1989}. This possibility was individually investigated for 3C~273 \citep{PaltaniTurler2003} and for 3C~454.3 \citep{FinkeDermer2010}. Other possible origin is the interaction between the jet and a red giant star stripping its outer shell to enrich the zone \citep{Bosch_Ramon2012,Khangulyan2013}. The same result can be obtained with a passing star-forming region interacting with the jet \citep{Zacharias2019}.

The highest levels of the Mg~II $\lambda$2798~\AA\ emission line and the UV Fe~II band fluxes and the 2010 gamma-ray outburst happen at the time when a jet component passes through or is ejected from the radio core of 3C~454.3. Similar behavior was observed for CTA~102 but in greater scale  \citep{Chavushyan2020}, the ratio between the maximum and minimum of $\lambda$3000~\AA\ continuum flux was $\sim179$, and for the Mg~II emission line and the UV Fe~II fluxes, the ratios were $\sim8$ and $\sim34$, respectively. 

In this study, we analyze the variability of the gamma-rays, X-rays, V-band, J-band, 1 mm, and 15 GHz light curves, additionally to the flux of the broad emission lines, specifically the Mg~II $\lambda$2798~\AA\ emission line and the UV Fe~II band from the optical spectra, as well as the $\lambda$3000~\AA\ continuum. The main goal of this work is to investigate the behaviors of these features to understand the physical conditions and processes that occurred during the high activity periods of 3C~454.3. 

The cosmological parameters adopted throughout this paper are H$_0=71$ km s$^{-1}$ Mpc$^{-1}$, $\Omega_{\Lambda}=0.73$, $\Omega_m=0.27$. At the redshift of the source, z=0.859, the spatial scale of 1\arcsec\ corresponds to a physical scale of 7.7 kpc and the luminosity distance is 5.489 Gpc.

\section{Observations}\label{sec:observe}

The Ground-based Observational Support of the Fermi Gamma-Ray Space Telescope at the University of Arizona has been monitoring 3C~454.3 throughout the years as a part of its program. Taking advantage of this great collection of data, in this work, we used 529 optical spectra observed from 2009-2018 calibrated against the V-band magnitude; see \cite{Smith2009} for further details of the observational setup and the data reduction process. The spectra were taken to the rest frame of the object and a cosmological correction of the form $(1+z)^3$ was applied to the flux \citep[e.g.][]{Peterson1997}. Galactic reddening correction was not applied since we are only interested in the relative flux changes. An important task in this work was to measure the flux of Mg~II $\lambda$2798~\AA , to achieve this we had to perform the subtraction of the UV Fe~II emission band and the continuum. The process of spectral fitting was done using the \texttt{SPECFIT} task from the \texttt{IRAF} package. 

\begin{figure}[htbp]
\begin{center}
\includegraphics[width=0.48\textwidth]{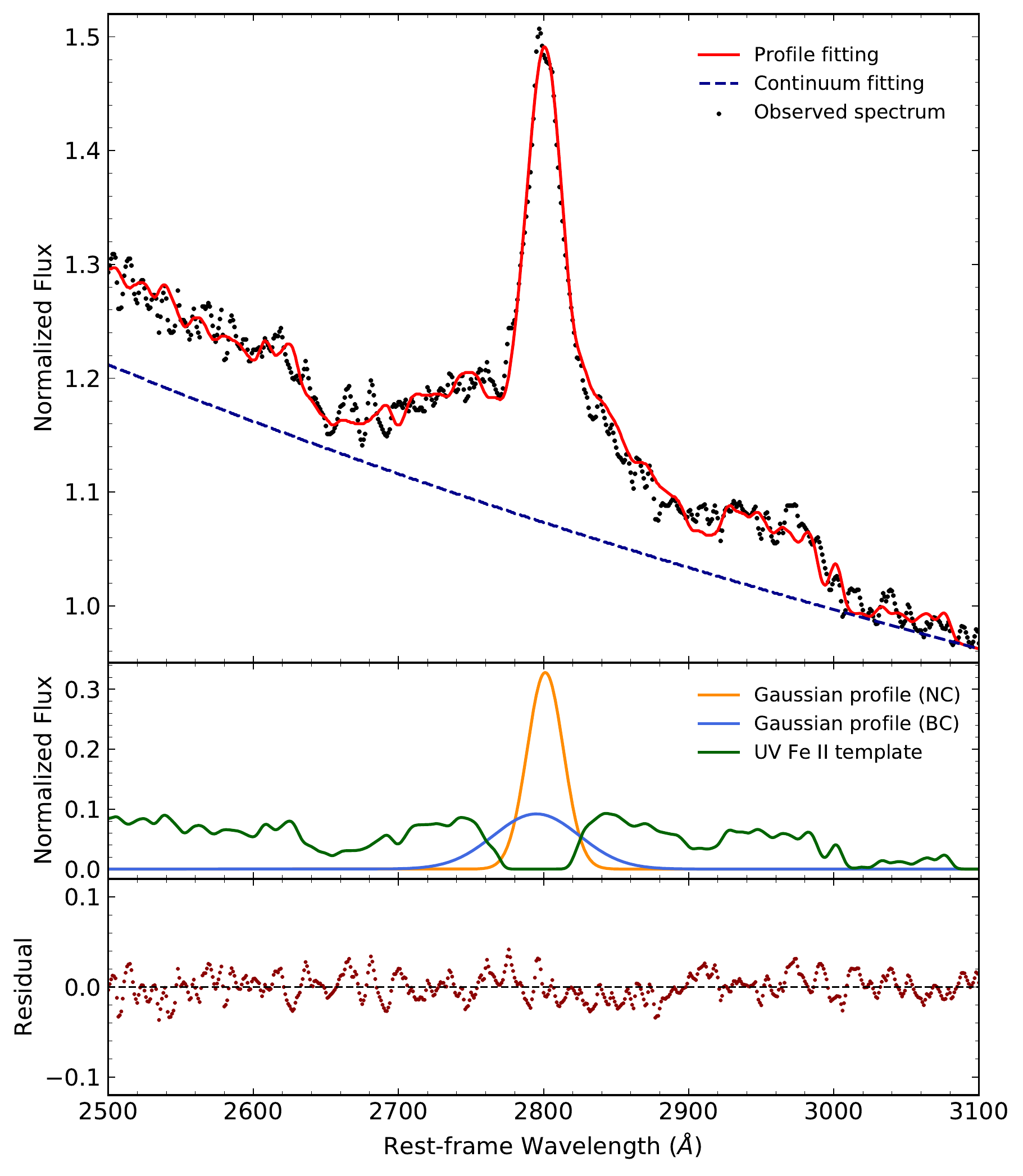}
\caption{Example decomposition of the Mg~II $\lambda$2798~\AA\ emission line and the UV~Fe~II band in a low continuum state spectrum taken at the Steward Observatory on May 10th, 2013. Top panel: Observed spectrum with the obtained best fit, a power-law function as the continuum is depicted. Middle panel: The broad and narrow components used to fit the Mg~II $\lambda$2798~\AA, as well as, the UV Fe II template. Bottom panel: The residuals from the subtraction of the best fit to the observed spectrum.}
\label{spec-decomp}
\end{center}
\end{figure}

\begin{table*}[htbp]
\centering
\caption{\tablenotemark{*}Sample of the flux measurements for the $\lambda$3000 \AA\ continuum, the Mg~II $\lambda$2798 \AA\ emission line and the UV Fe II band.}
\begin{tabular}{ccccccc}
\hline
\hline
\multirow{2}{*}{JD-2450000} & {Flux Continuum $\lambda$3000 \AA\ } &  {Flux Mg~II $\lambda$2798 \AA\ } &  {Flux UV Fe II }  \\
 & $\times10^{-13}\, erg\, s^{-1}cm^{-2}\, {\rm \AA}^{-1}$ & $\times10^{-13}\, erg\, s^{-1}cm^{-2}$ & $\times10^{-13}\, erg\, s^{-1}\, cm^{-2}$ \\
\hline
5086.84 &    0.200 $\pm$    0.004 &    1.26 $\pm$    0.14 &    0.56 $\pm$    0.17 \\
5088.76 &    0.235 $\pm$    0.004 &    1.28 $\pm$    0.14 &    0.85 $\pm$    0.17 \\
5088.90 &    0.215 $\pm$    0.004 &    1.22 $\pm$    0.13 &    0.76 $\pm$    0.16 \\
5089.68 &    0.186 $\pm$    0.004 &    1.23 $\pm$    0.13 &    0.66 $\pm$    0.16 \\
5089.81 &    0.184 $\pm$    0.003 &    1.23 $\pm$    0.11 &    0.64 $\pm$    0.13 \\
5089.94 &    0.184 $\pm$    0.004 &    1.19 $\pm$    0.15 &    0.66 $\pm$    0.19 \\
5090.68 &    0.194 $\pm$    0.003 &    1.26 $\pm$    0.09 &    0.66 $\pm$    0.11 \\
5090.81 &    0.197 $\pm$    0.004 &    1.17 $\pm$    0.14 &    0.54 $\pm$    0.17 \\
5090.94 &    0.204 $\pm$    0.004 &    1.22 $\pm$    0.14 &    0.71 $\pm$    0.18 \\
5091.66 &    0.267 $\pm$    0.005 &    1.13 $\pm$    0.17 &    0.73 $\pm$    0.20 \\
\hline
\end{tabular}
\tablenotetext{*}{The complete table is available in a machine-readable form in the online journal.}
\label{Table_flux}
\end{table*}

First, the continuum was approximated as a power-law function to subtract it from the spectrum. Next, the Fe~II emission was also subtracted by fitting it using the template of \cite{Vestergaard2001}. After performing this process for all the spectra, we integrated the Mg~II line profile in the range of $2725-2875$~\AA\ to obtain its flux. We measured the Fe~II flux by integrating the spectrum free of the continuum and the Mg~II contribution in the range of $2850-3000$~\AA . A third feature obtained from the spectra was the flux of the UV-continuum at 3000~\AA\, which consisted of taking the average value in the range $2950-3050$~\AA\ from the iron-subtracted spectrum. An example of the spectral decomposition is illustrated in Figure \ref{spec-decomp}. Unfortunately, we were not able to perform an analysis of the Mg~II emission line FWHM evolution, since the spectra were taken with different slit sizes (3\arcsec being the smallest and 7\farcs6 the largest).

There are two unique contributions to the corresponding uncertainty in the Mg~II flux measurement. The first and main contribution is the random error, in this case, produced by the dispersion and the signal-to-noise ratio of the spectra, which was estimated as in \cite{Tresse1999}. The second contribution is introduced by the subtraction of the Fe~II emission, estimated as in \cite{Leon-Tavares2013}, considering that in the range $2786-2805$ \AA\ no iron subtraction was performed.
For the corresponding error in the Fe~II flux, the only contribution is the one due to the random error, estimated similarly as for Mg~II. Lastly, the error for the estimation of the $\lambda$3000~\AA\ continuum flux was calculated taking the root mean square of the iron-subtracted spectrum around $\lambda$3000$\pm$50~\AA.  For further details, see \cite{Leon-Tavares2013,Chavushyan2020}. A sample of the measured fluxes and errors is displayed in Table \ref{Table_flux}. The complete table is available in the online journal in machine-readable format.

\begin{figure*}[htbp]
\begin{center}
\includegraphics[width=1\textwidth]{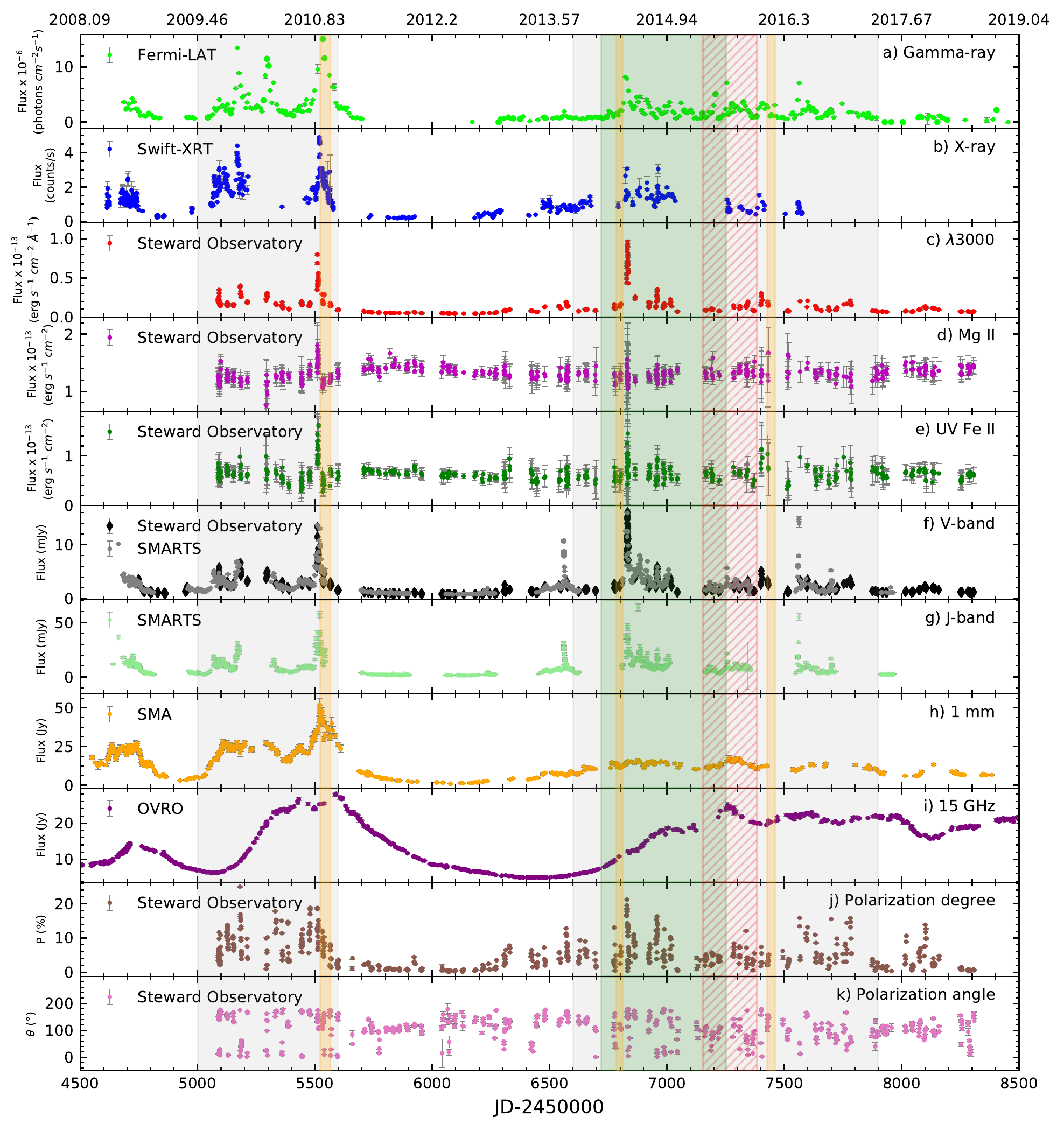}
\caption{Multiwavelength light curves for 3C~454.3 (see Section \ref{sec:observe} for details on the observations).  The first and second orange vertical stripes show the time when the knots B11 \citep{Jorstad2017} and K14 \citep{Liodakis2020} were ejected from the radio core, respectively. The third orange vertical stripe represents the time at which the knot K16 \citep{Weaver2019} crossed the 43 GHz core. The green vertical stripe represents the expected time for which B11 would have passed through the quasi-stationary component \citep[C,][]{Jorstad2013,Jorstad2017}. The width of these stripes represent their respective uncertainty. The red hatched vertical stripe shows the duration of the flare in the 43 GHz flux density of the quasi-stationary component C taken from Figure 7 of \cite{Liodakis2020}. The grey vertical stripes represent the flaring periods as defined in subsection \ref{cc_flares}.}
\label{multiwavelength}
\end{center}
\end{figure*}

The gamma-ray light curve was built using the data from the public database of the \textit{Fermi} Large Area Telescope \citep[LAT,][]{Abdo2009} in the energy range $0.1-300$ GeV with the \texttt{Fermitools version 1.0.2}. All sources within 15 degrees of the location of 3C~454.3 were included in the model, taken from the 4FGL catalog \citep{4FGL}. 
The X-ray data were obtained from the public database of the Swift X-Ray Telescope (XRT). The Swift-XRT data were processed using the standard SWIFT tools (Swift Software version 3.9, \texttt{FTOOLS version 6.12} and \texttt{XSPEC version 12.7.1}) and the light curves were generated with \texttt{xrtgrblc version 1.6}. Details of the reduction procedure can be found in \cite{Stroh2013}.
We took the optical V-band data from two sources, the Ground-based Observational Support of the Fermi Gamma-Ray Space Telescope at the University of Arizona \citep[Steward Observatory,][]{Smith2009} and the Small and Moderate Aperture Research Telescope System \citep[SMARTS,][]{Bonning2012}.
The Near-Infrared (NIR) J-band data were obtained from SMARTS as well.
The 1~mm data were obtained from the Sub-Millimeter Array (SMA) public database. The observations and data reduction details can be found in \cite{Gurwell2007}.
We used the 15~GHz data from the Owens Valley Radio Observatory \citep[OVRO,][]{Richards2011} public database.
The optical linear polarization degree and the optical polarization angle were retrieved from the Steward Observatory. The light curves of all the described data are shown in Figure \ref{multiwavelength}.

\section{Cross-correlation analysis} \label{sec:crosscorr}

\subsection{Cross-correlations for the complete dataset}\label{cc_complete}

Cross-correlation analysis was performed to determine the time delay between the emission of the different bands. Only the time delays with a correlation coefficient above the 99\% significance level were considered. Three different methods were used to ensure the truthfulness of the results, following the methodology in \cite{PatinoAlvarez2018}. Furthermore, an alias check via Fourier analysis was conducted to identify and remove these from the results \citep{Press2007}\footnote{\url{https://www.cambridge.org/numericalrecipes}}. After the alias check, some cross-correlation analyses showed more than one time delay. These were included in the results as well. From visual inspection in Figure \ref{multiwavelength}, we notice that there is a time delay consistent with zero between the gamma-rays, V-band, J-band, and the $\lambda3000$\AA\ continuum emission, which was confirmed through the cross-correlation analysis. This result indicates that the emission regions of the continuum, the V-band, and the J-band, as well as the source of the seed photons responsible for the IC scattering, are co-spatial. Furthermore, the Mg~II $\lambda2798$\AA\ emission line presents anti-correlations against the gamma-rays at $279.9\pm8.8$ days, the $\lambda3000$\AA\ continuum at $19.8\pm4.2$ days (continuum vs Mg~II), and the Fe~II at $-15.9\pm4.2$ days. The Mg~II presents a significant correlation with the $\lambda3000$\AA\ continuum at a time delay of $637.6\pm4.2$ days (continuum vs Mg~II), which is consistent with the value ($621\pm45$ days) reported by \cite{Nalewajko2019}. The UV Fe~II  band presents correlations at a time delay consistent with zero against the continuum, gamma-rays, and Mg~II emission, however, there is a correlation with the continuum at a time delay of $576.9\pm4.2$ days (continuum vs Fe~II). Moreover, the Fe~II also presents anti-correlations; the one mentioned above with Mg~II and with the continuum at a time delay of $20.4\pm4.2$ days (continuum vs Fe~II). The full cross-correlation results are shown in Table \ref{Table_CC}. Some of the resulting figures from the cross-correlations are displayed in the Figure \ref{cc-fig}. The remaining figures are available in the online journal.

\begin{table*}
\centering
\caption{Cross-correlation results for the full light curves given in time delays (days) with their uncertainty at 90\% confidence level. All delays have correlations at the $\geq$ 99\% significance level. The Delay 2 and Delay 3 are additional delays found to be statistically significant, and only shown for the cross-correlation analyses in which the delays are not aliases. All cross-correlations are performed in the order stated in this table.}

\begin{tabular}{lccc}
\hline
\hline
                Bands &         Delay 1                       & Delay 2           & Delay 3        \\
\hline
        1mm versus 3000\AA &           -18.6 $\pm$              16.6 &     ... &     ...       \\
          1mm versus Fe II &            -6.3 $\pm$               4.7 &     ... &     ...       \\
     1mm versus Gamma-rays &            -0.1 $\pm$               8.8 &     ... &     ...       \\
              1mm versus J &            -8.2 $\pm$               4.7 &     ... &     ...       \\
          1mm versus Mg II &  Not conclusive                       &     ... &     ...       \\
         1mm versus X-rays &            -1.5 $\pm$               5.3 &     ... &     ...       \\
         15 GHz versus 1mm &          -131.6 $\pm$              26.5 &     ... &     ...       \\
     15 GHz versus 3000\AA &  Not conclusive                       &     ... &     ...       \\
       15 GHz versus Fe II &  No correlation                       &     ... &     ...       \\
  15 GHz versus Gamma-rays &  No correlation                       &     ... &     ...       \\
           15 GHz versus J &          -157.2 $\pm$               2.9 &     ... &     ...       \\
       15 GHz versus Mg II &  Not conclusive                       &     ... &     ...       \\
      15 GHz versus X-rays &          -137.5 $\pm$               5.3 &     ... &     ...       \\
     3000\AA\ versus Fe II &             0.0 $\pm$               4.2 &   576.9 $\pm$   4.2 &   20.4  $\pm$  4.2* \\
3000\AA\ versus Gamma-rays &            -1.5 $\pm$               8.8 &     ... &     ...       \\
     3000\AA\ versus Mg II &           637.6 $\pm$               4.2 &   19.8  $\pm$  4.2* &  ... \\
    3000\AA\ versus X-rays &             4.1 $\pm$               5.3 &   231.5 $\pm$   5.3 &  ... \\
   Mg II versus Gamma-rays &  No correlation                       &     ... &     ...       \\
        Mg II versus Fe II &             2.5 $\pm$               4.2 &  -15.9  $\pm$  4.2* &  ... \\
       Mg II versus X-rays &          -24.4  $\pm$             28.3* &     ... &     ...       \\
   Fe II versus Gamma-rays &            -0.5 $\pm$               8.8 &     ... &     ...       \\
       Fe II versus X-rays &             7.2 $\pm$               5.3 &     ... &     ...       \\
          J versus 3000\AA &             1.5 $\pm$               4.2 &     ... &     ...       \\
            J versus Fe II &            2.25 $\pm$               4.2 &     ... &     ...       \\
       J versus Gamma-rays &           -0.57 $\pm$               8.8 &     ... &     ...       \\
            J versus Mg II &           24.2  $\pm$              4.2* &     ... &     ...       \\
           J versus X-rays &             1.3 $\pm$               5.3 &     ... &     ...       \\
              V versus 1mm &            10.4 $\pm$               4.7 &     ... &     ...       \\
           V versus 15 GHz &           150.0  $^{+48.39}_{-39.54}$ &     ... &     ...       \\
         V versus 3000\AA\ &             1.6 $\pm$               4.2 &     ... &     ...       \\
            V versus Fe II &             1.6 $\pm$               4.2 &     ... &     ...       \\
       V versus Gamma-rays &             2.1 $\pm$               8.8 &     ... &     ...       \\
                V versus J &             0.1 $\pm$               2.8 &     ... &     ...       \\
            V versus Mg II &           21.3  $\pm$              4.2* &     ... &     ...       \\
           V versus X-rays &             3.9 $\pm$               5.8 &     ... &     ...       \\
       V versus Gamma-rays &             2.1 $\pm$               8.8 &     ... &     ...       \\
  X-rays versus Gamma-rays &             4.5 $\pm$               8.8 &     ... &     ...       \\
\hline
\end{tabular}

\label{Table_CC}
\tablenotetext{*}{~Anti-correlation at this delay.}
\end{table*}

\subsection{Cross-correlations for the separate flaring periods}\label{cc_flares}

Aiming to understand some of the time delays found in the cross-correlations done for the complete data set, and with the purpose to find the dominant gamma-ray emission mechanism for each flaring period; we selected from the light curves (Figure \ref{multiwavelength}), considering the high activity states in the UV-continuum, two sets of observations. The period of $5000 < \text{JD}-2450000 < 5600$ (flaring period 1 - FP1) and the period of $6600 < \text{JD}-2450000 < 7900$ (flaring period 2 - FP2). These flaring periods are marked in Figure~\ref{multiwavelength} with grey vertical stripes. Then, we performed the cross-correlations between the gamma-rays and V-band, as well as, the Mg~II, UV~Fe~II, and $\lambda$3000\AA\ continuum. For the FP1, we did not find anti-correlation between any of the light curves. The results show a time delay not consistent with zero for gamma-rays against the UV-continuum and the Fe~II. For the FP2, we found an anti-correlation between the Mg~II and the UV-continuum at $44.64\pm4.59$ days. This time the gamma-rays correlate with the UV-continuum and Fe~II at a time delay consistent with zero. We found that the Fe~II correlates with the UV-continuum at a time delay consistent with zero in both periods. It is important to note that the Mg~II does not correlate with the gamma-rays in either of the periods. The cross-correlation results for this analysis are shown in Table \ref{Table_CCflares}.
Some of the resulting figures from the cross-correlations for the FP1 and FP2 are displayed in the Figures \ref{cc-figfp1} and \ref{cc-figfp2}, respectively. The remaining figures are available in the online journal.

\begin{table*}
\centering
\caption{Cross-correlation results for the separate flaring periods, as described in subsection \ref{cc_flares}, given in time delays (days) with their uncertainty at 90\% confidence level. All delays have correlations at the $\geq$ 99\% significance level. The Delay 2 is an additional delay found to be statistically significant and only shown for the cross-correlation analyses in which the delay is not an alias. All cross-correlations are performed in the order stated in this table.}
\begin{tabular}{lccc}
\hline
\hline
                       &                                      \multicolumn2c{FP1} &                           FP2 \\
                 Bands &                      Delay 1 &                   Delay 2 &                        Delay  \\
\hline
      Mg II versus 3000\AA &            3.8 $\pm$      2.4 &       ...                 &          44.6 $\pm$     4.6* \\
      Fe II versus 3000\AA &            0.0 $\pm$      2.4 &       ...                 &            0.1 $\pm$      4.6 \\
        Fe II versus Mg II &           -2.7 $\pm$      2.4 &    -36.8 $\pm$        2.4 &  No correlation               \\
 Gamma-rays versus 3000\AA &           -6.6 $\pm$      2.4 &       ...                 &            3.0 $\pm$      4.6 \\
   Gamma-rays versus Fe II &           -6.4 $\pm$      2.4 &    -36.4 $\pm$        2.4 &            2.6 $\pm$      4.6 \\
   Gamma-rays versus Mg II &  No correlation               &       ...                 &  No correlation               \\
         V versus 3000\AA\ &            0.4 $\pm$      2.4 &       ...                 &            0.4 $\pm$      4.6 \\
            V versus Mg II &           -1.2 $\pm$      2.4 &       ...                 &  No correlation               \\
            V versus Fe II &           -0.2 $\pm$      2.4 &       ...                 &            0.7 $\pm$      4.6 \\
       V versus Gamma-rays &            3.9 $\pm$      1.5 &       ...                 &            0.6 $\pm$      2.0 \\
\hline
\end{tabular}
\tablenotetext{*}{~Anti-correlation at this delay.}
\label{Table_CCflares}
\end{table*}

\section{Variability}\label{sec:var}

\begin{figure*}[htbp]
\begin{center}
\includegraphics[width=1\textwidth]{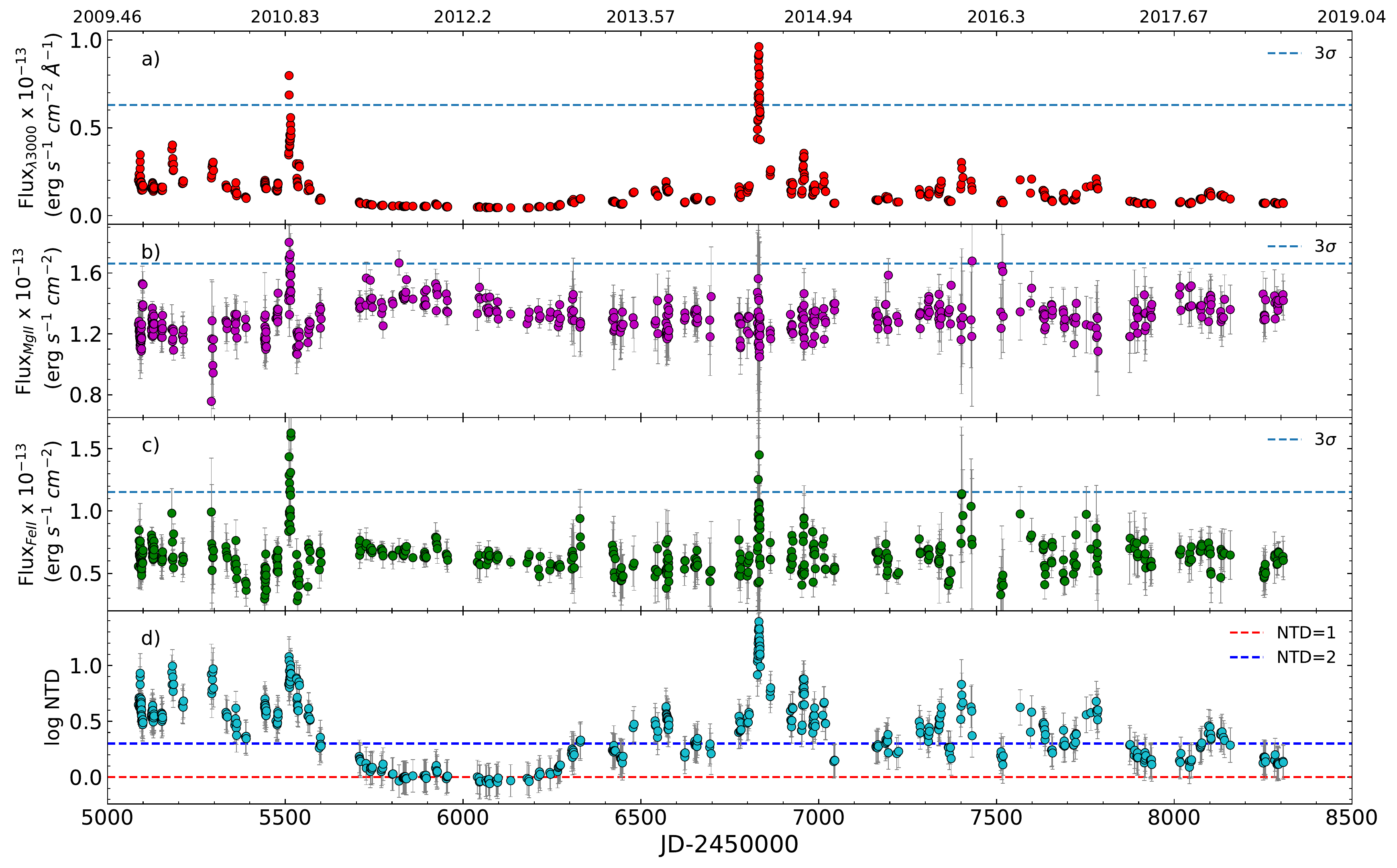}
\caption{Light curves of a) the $\lambda$3000 \AA\ continuum flux, b) the Mg~II $\lambda$2798 \AA\ emission line flux, c) the UV Fe~II band flux, and d) the NTD parameter in logarithmic scale. In the first three panels, the blue dashed line represents $3\sigma$. The red and blue dashed lines in d) represent NTD=1 and NTD=2, respectively. }
\label{line-continuum}
\end{center}
\end{figure*}

For the two most significant UV-continuum flare-like events, 2010 (maximum value in $\text{JD}-2450000=5510.58$) and 2014 (maximum value in $\text{JD}-2450000=6831.96$), the Mg~II emission line and the UV~Fe~II band response was different. On the one hand, during the 2010 event, 3C~454.3 had a major gamma-ray outburst which had its counterpart in the UV-continuum. The Mg~II and the Fe~II both significantly responded to this increase in the ionizing continuum, corresponding with previous results of \cite{Leon-Tavares2013}, \cite{Isler2013}, and more recently \cite{Nalewajko2019}.  On the other hand, during the 2014 event, the source presented a gamma-ray outburst smaller than the one in 2010 but the increase in the UV-continuum flux was higher. The Mg~II flux did not significantly increase in response to the continuum, a similar behavior was reported by  \cite{Nalewajko2019}. However, the UV Fe~II presents two flux points in their light curve above 3$\sigma$ which suggests that Fe~II responded to this flaring event. These results can be seen in Figure \ref{line-continuum}.

Another tool that can help to understand what was happening during these major events is the Non-Thermal Dominance parameter \citep[NTD,][]{Shaw2012,PatinoAlvarez2016,Chavushyan2020}. The calculation of the NTD, as its name suggests, constitutes a method to find out if there is a non-thermal contribution (the jet as the source) to the continuum emission. If the NTD~$=1$, the continuum emission is purely due to thermal emission, specifically, the source of the emission being the accretion disk. If the NTD lies between 1 and 2 then, the dominant source of the continuum is the accretion disk, but some contribution from the jet is expected. The non-thermal emission dominates for values of NTD~$>2$, and if NTD~$=2$ the contribution from the disk and the jet is equal. For the calculation of this parameter, one needs to know the observed continuum luminosity ($L_{\text{obs}}$) and estimate from the emission line luminosity a predicted continuum luminosity ($L_{\text{pred}}$, expected luminosity of the accretion disk), NTD~$=L_{\text{obs}}/L_{\text{pred}}$. We calculated this predicted continuum luminosity from the bisector fit of the relation between the luminosity of Mg~II and the $\lambda3000$\AA\ continuum luminosity found by \cite{Shen2011}. In Figure \ref{line-continuum} panel d), we can see that during the high energy events the NTD value surpasses NTD~$=2$ and, during the 2010 and 2014 flare the NTD reaches the maximum values of 12.0 and 24.7, respectively. This result indicates that the continuum emission during these flares is vastly dominated by the jet. In Figure \ref{luminosities} panel c), we can see that NTD increases when the continuum emission luminosity increases, which is expected. Furthermore, just after the 2010 flare ($5700 < \text{JD}-2450000 < 6300$) the NTD value oscillates around 1 (the disk as the dominant source of the continuum), which coincides with the gamma-rays (emitted by the jet) not being detected.

The linear optical polarization is a tool for probing the magnetic field in the jet \citep[e.g.][and references therein]{Zhang2019}. From the observations presented in Figure~\ref{multiwavelength} panels i) and j), we can determine that the optical linear degree of polarization reaches its highest values at the time of the major gamma-ray outbursts. Indicating that either a shock is compressing an initially turbulent magnetic field \citep{Laing1980} or a strongly polarized emission zone emerged at the time \citep{Itoh2013}. Equally notable are the large changes in the polarization angle during the high activity periods, this tells us that the magnetic field of the jet is changing its morphology and even the structure of the jet itself \citep{Marscher2008,Chandra2015,Blinov2018}. 

At the end of 2010, a blob \citep[B11,][]{Jorstad2017} was ejected from the 43 GHz core of 3C~454.3 ($\text{JD}-2450000=5544.25\pm21.90$) causing the brightest gamma-ray flare of that year (maximum value in $\text{JD}-2450000=5512.66$).  This ejection is represented by an orange stripe in Figure \ref{multiwavelength}. The deprojected distance from the black hole to the  43 GHz core, estimated with an angle of 1\fdg3 (between the jet and our line of sight), is $\sim$9~pc \citep{Kutkin2014}. Meanwhile, \cite{Jorstad2010} estimated an upper limit of $\sim$18~pc. In this work we will adopt $\sim$9~pc as deprojected distance from the black hole to the 43 GHz core.
Using the mean proper motion ($0.152\pm0.012$ mas/yr) of the blob B11 given by \cite{Jorstad2017} calculated with 43 GHz VLBI observations, we estimated the time for which it would reach the quasi-stationary component C \citep[at $\sim$0.6~mas from the radio core which corresponds to a projected $\sim$4.6~pc,][]{Jorstad2013,Jorstad2017}. The expected time for which the blob B11 would have reached the quasi-stationary component C should be $6719.46\leq\text{JD}-2450000\leq 7252.65$ (green vertical stripe in Figure \ref{multiwavelength}). These quasi-stationary features are typically interpreted as recollimation shocks in an overpressured jet and cannot be explained by traveling shock waves within a pressure matched and, therefore, conical jet \citep{DalyMarscher1988}.

Recently, it was reported the ejection of the relativistic component K14 from the 43 GHz core ($\text{JD}-2450000=6797\pm15$) which coincides with the maximum of the gamma-ray flare of 2014 (maximum value in $\text{JD}-2450000=6821.66$) within $2\sigma$ of uncertainty \citep{Liodakis2020}. This ejection is illustrated in Figure \ref{multiwavelength} with an orange stripe. Additionally, in the bottom panel of Figure 7 of their work, the quasi-stationary component C increased its 43 GHz flux density $\sim$9 times during the FP2 (maximum value in $\text{JD}-2450000\sim7287$), which we represent with a red hatched vertical stripe in Figure \ref{multiwavelength} (the width represents the time from the beginning of the flare until the end of the data in the Figure 7 of the authors). We observe this flaring activity in gamma-rays, V-band and J-band (maximum values in $\text{JD}-2450000=7255.66,\ 7256.68$ and $7256.68$, respectively) as well. We can see in Figure \ref{multiwavelength} that the green vertical stripe coincides within its uncertainty with the red hatched vertical stripe. Hence, the estimated collision between the blob B11 and the quasi-stationary component C could explain the origin of the multiwavelength flaring event in the latter. 

\cite{Weaver2019} reported an ejection of a relativistic component K16 from the 43 GHz core $\sim$4 months before the flaring event of June 2016 ($\text{JD}-2450000=7448\pm17.5$) and traveled across until it disappeared. They interpreted that the electrons were accelerated from a back interaction with the core causing the variability. This ejection is marked with an orange stripe in Figure \ref{multiwavelength}.

\begin{figure*}[htbp]
\begin{center}
\includegraphics[width=1\textwidth]{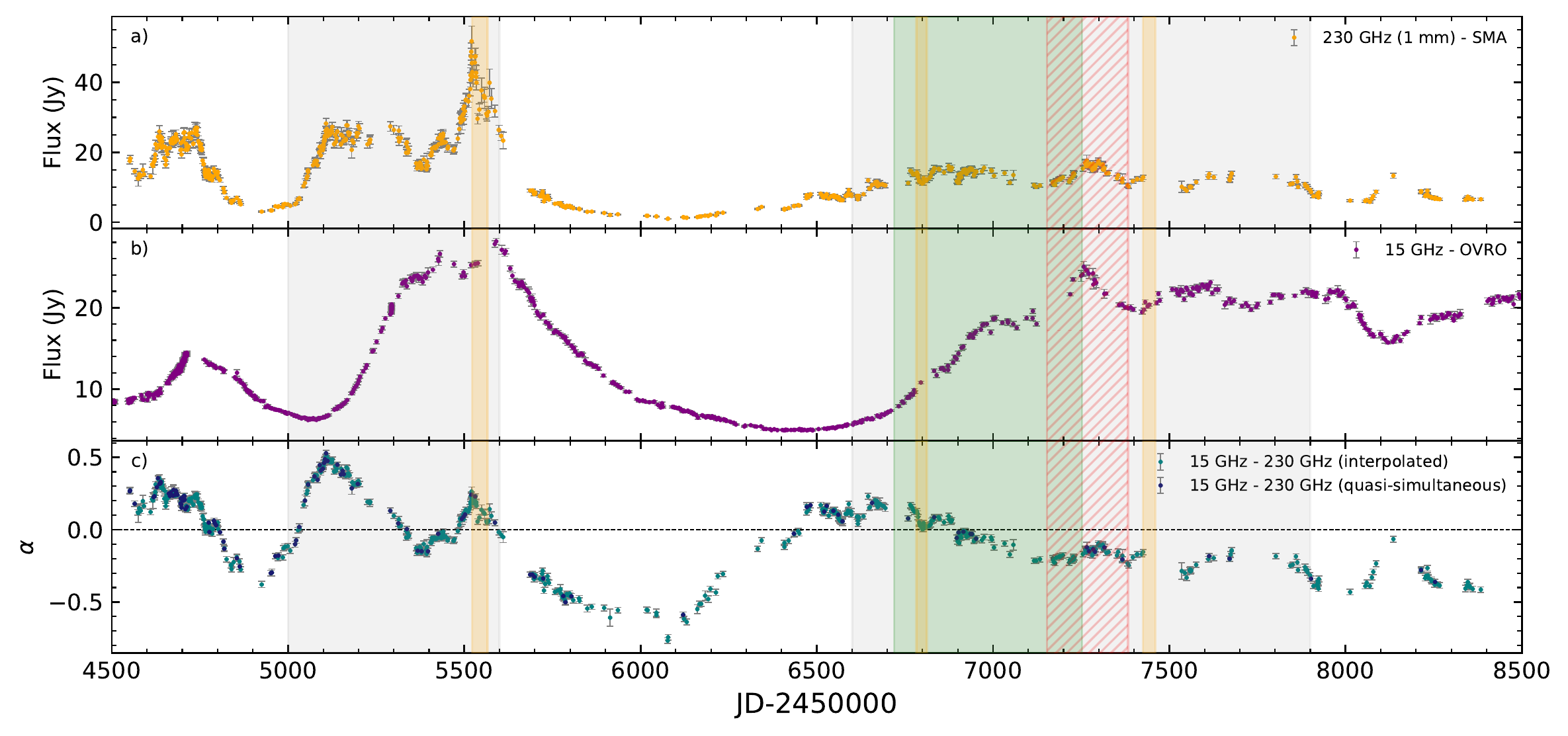}
\caption{Light curves of a) 1 mm from the public database of the SMA and b) 15 GHz from the OVRO. c) The spectral index ($\alpha$, $S_\nu \propto \nu^\alpha$) estimated with the 15 GHz and 1 mm (we adopted 230 GHz) flux densities. The teal points correspond to the spectral indices estimated through interpolation and the dark blue points correspond to the spectral indices estimated only on quasi-simultaneous epochs. The horizontal dashed line is at $\alpha = 0$. The vertical stripes represent the same as in Figure \ref{multiwavelength}.}
\label{spectral_index}
\end{center}
\end{figure*}

There are noticeable increases in the fluxes of 1 mm and 15 GHz during the high activity periods of the object in Figure \ref{multiwavelength}. These increases indicate that the main source responsible for the flaring events is the jet (non-thermal emission). The 15 GHz increased its flux in a comparable amount during the two flaring periods. On the other hand, we see that for the FP2, the 1 mm flux increases less than during the FP1. A similar behavior can be seen for the 22 GHz and 37 GHz light curves in Figure 1 from \cite{Sarkar2019}. In an attempt to explain this effect, we estimated the spectral index ($\alpha$, $S_\nu \propto \nu^\alpha$) with the 15 GHz and 1 mm (adopting 230 GHz as the frequency) flux densities during the quasi-simultaneous epochs and interpolating the 15 GHz light curve into the 1 mm epochs. Both estimates are displayed in Figure \ref{spectral_index}. In the panel c) of this figure, the dashed line at $\alpha = 0$ \citep{Fromm2011,ParkTrippe2014} indicates the separation between optically thin ($\alpha < 0$) and optically thick ($\alpha > 0$). During the 2010 flare, the spectral index indicates that the spectrum is in the optically thick regime. However, by the time of the brightest flare of 2014, the spectral index is very close to the threshold and after $JD-2450000 \approx 6900$, the spectral index goes below zero (optically thin), letting us see a larger region. Furthermore, the lower energy electrons (e.g. 15 GHz) are cooled by synchrotron less efficiently than higher energy electrons (e.g. 1 mm), making the more energetic electrons to stay confined in a smaller region. The combination of these factors could be the reason of why the 1 mm flux had a smaller response during the second flaring period in comparison to the first.

The variability of the spectral index indicates variations of intricate source properties, most likely changes in the optical depth \citep{Trippe2011}. Along the same lines, \cite{Park2019} interpreted that when the radio spectra are optically thick an emerging component is close to the core and when the spectra become optically thin the component is well separated from the core. This interpretation describes well what happened during the FP1 and is compatible with the shock-in-jet model \citep[e.g.][]{MarscherGear1985,Valtaoja1992,Fromm2011,Hughes2011,Weaver2019}.
During the FP2, the spectral index is not varying as much as during the FP1 and the changes are slower. This could indicate that the spectra are different during these periods, hence, the physical conditions and/or mechanism as well.
The more complex nature of the FP2, as a result of a superposition of multiple events, might justify the strange behavior of the spectral index. Finally, since there is a flaring behavior of the quasi-stationary component C, in addition to the multiple ejections from the radio core during the FP2, we can hypothesize that there are multiple gamma-ray emission regions in the jet of 3C~454.3.

\section{Luminosity relations}\label{sec:LumRel}

\subsection{Luminosity relations for the complete dataset}\label{Lum_complete}

\begin{figure}[htbp]
\begin{center}
\includegraphics[width=0.47\textwidth]{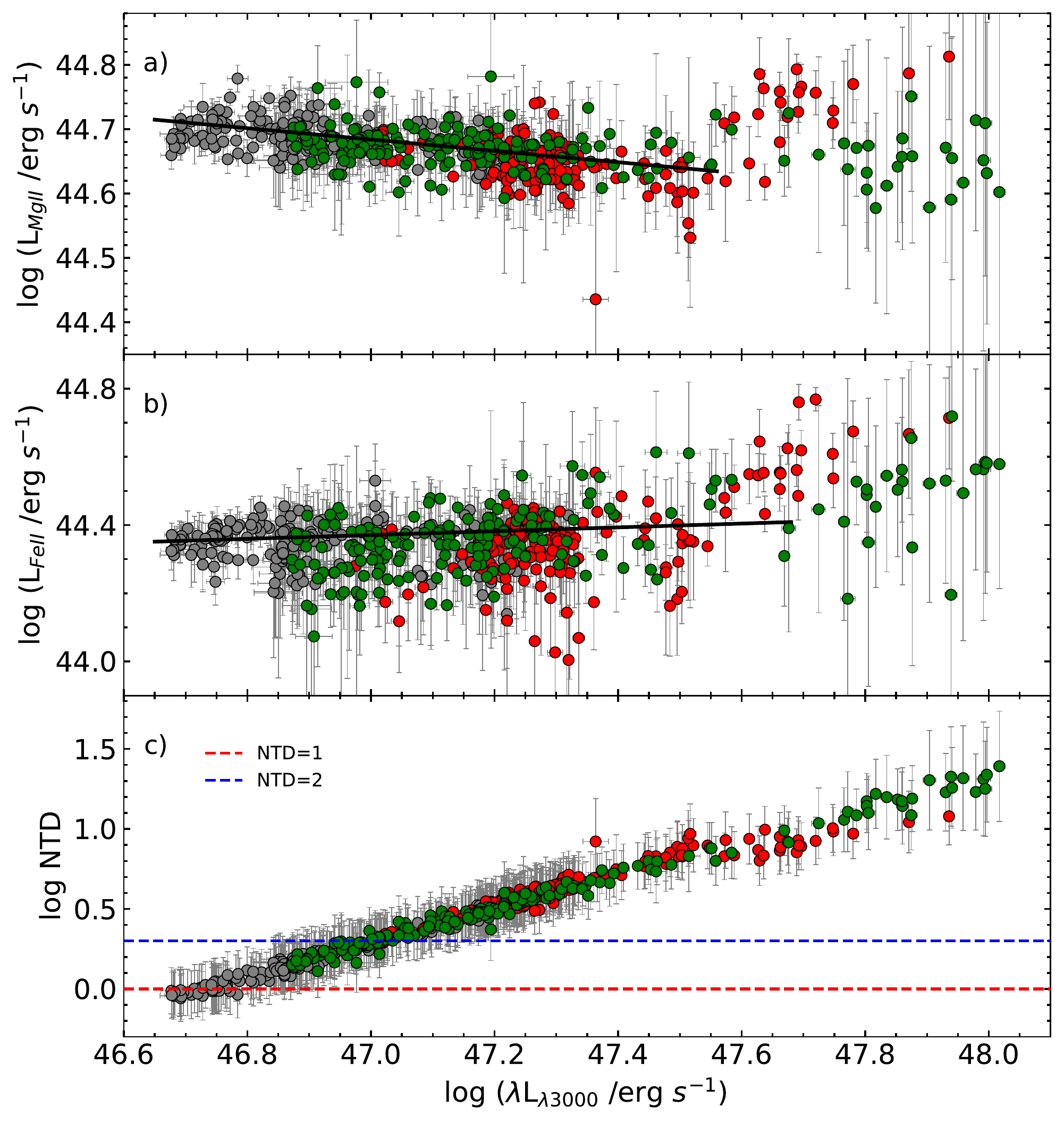}
\caption{Variations of $\lambda$3000 \AA\ continuum luminosity compared to a) the Mg~II $\lambda$2798 \AA\ emission-line 
luminosity. b) The UV Fe~II band luminosity. c) The NTD parameter in the logarithmic scale; NTD~$=1$ and NTD~$=2$ are represented by red and blue dashed lines, respectively. The last statistically significant linear fit (see Section \ref{Lum_complete}) is represented by the black line in panels a) and b). The colors indicate the period of observation: the green corresponds to the observations during the FP1 (see Section \ref{cc_flares}), the red to the observations during the FP2 (see Section \ref{cc_flares}), and the grey to the remaining observing points.}
\label{luminosities}
\end{center}
\end{figure}

With the purpose of directly compare the response of the Mg~II and Fe~II to the $\lambda3000$ \AA\ continuum, we calculated their luminosities and plotted them as shown in Figure \ref{luminosities}. We performed linear fitting for each luminosity relation using Orthogonal Distance Regression from the \texttt{SciPy ODR package}\footnote{\url{https://docs.scipy.org/doc/scipy/reference/odr.html}}. A fit was considered statistically significant when the respective p-value ($p_v$, the probability of obtaining a chi-square value equal or higher by chance) is below 0.05.  In the panel a), we can see that the luminosity of the Mg~II emission line tends to decrease with the increase of the luminosity of the continuum, except for the highest luminosities. For the entire dataset, we found that the linear fit for this relation did not result statistically significant. To find the continuum luminosity up to where the luminosity of Mg~II is decreased, we looked for the linear fit for the first 11 points \citep[minimum number for a trustworthy correlation,][]{Alexander1997}, then for 12 and so forth in the direction of the increasing continuum luminosity. We found that the $p_v$ varied through all the analyses. The last significant linear fit (black line in the panel a of Figure \ref{luminosities}) was at $\log (\lambda L_{\lambda 3000}/$erg s$^{-1})=47.56$ giving a resulting slope of $-0.09\pm0.01$ with a $p_v = 0.0329$. Additionally, we performed a Spearman correlation rank test and found a significant anti-correlation ($\rho=-0.41$, $p_v=2\times 10^{-23}$). In the case of the Spearman correlation, the $p_v$ refers to the probability of obtaining a correlation coefficient equal or higher by chance.

For the Fe~II in Figure \ref{luminosities} panel b), we cannot see a clear behavior at first glance, hence, we performed the same analysis done for Mg~II. The last statistically significant linear fit (black line in the panel b of Figure \ref{luminosities}) was at $\log (\lambda L_{\lambda 3000}/$erg s$^{-1})=47.69$, giving a resulting slope of $0.06\pm0.01$ with a $p_v = 0.037$, above this threshold the fit becomes non-significant. The Spearman test shows that there is a weak correlation ($\rho=0.25$, $p_v=4\times 10^{-09}$), it is important to note that the $\rho^2$ is less than the error rate for almost all the points (any correlated variability has a lower amplitude than the uncertainty).

Since the object has two different phases, disk dominance, and jet dominance, we explored the same relations between the Mg~II and Fe~II luminosities against the ionizing continuum to find whether the disk or the jet is responsible for the above correlations. We separated the data into two sets, one containing the points when the NTD~$<2$ and the other of when NTD~$>2$. The NTD~$<2$ dataset contains 187 points corresponding to 35$\%$ of the total observation points, which means that in 65$\%$ of the observing time the jet is the dominant source of the continuum.  During the disk dominance phase, the Spearman test shows that there is no correlation between the continuum and the spectral features, except for a weak anti-correlation between the continuum and the Mg~II ($\rho=-0.20$, $p_v=0.004$). However, during the jet dominance phase, we found that there is a significant correlation between the continuum and the Fe~II luminosities ($\rho=0.39$, $p_v=7\times 10^{-14}$), and likewise with the ratio between Fe~II and Mg~II luminosities ($\rho=0.47$, $p_v=6\times 10^{-20}$). Again, between the continuum and Mg~II, there is a weak anti-correlation ($\rho=-0.15$, $p_v=0.005$).

\subsection{Luminosity relations for the separate periods of flares}\label{Lum_flares}

\begin{figure*}[htbp]
\begin{center}
\includegraphics[width=1\textwidth]{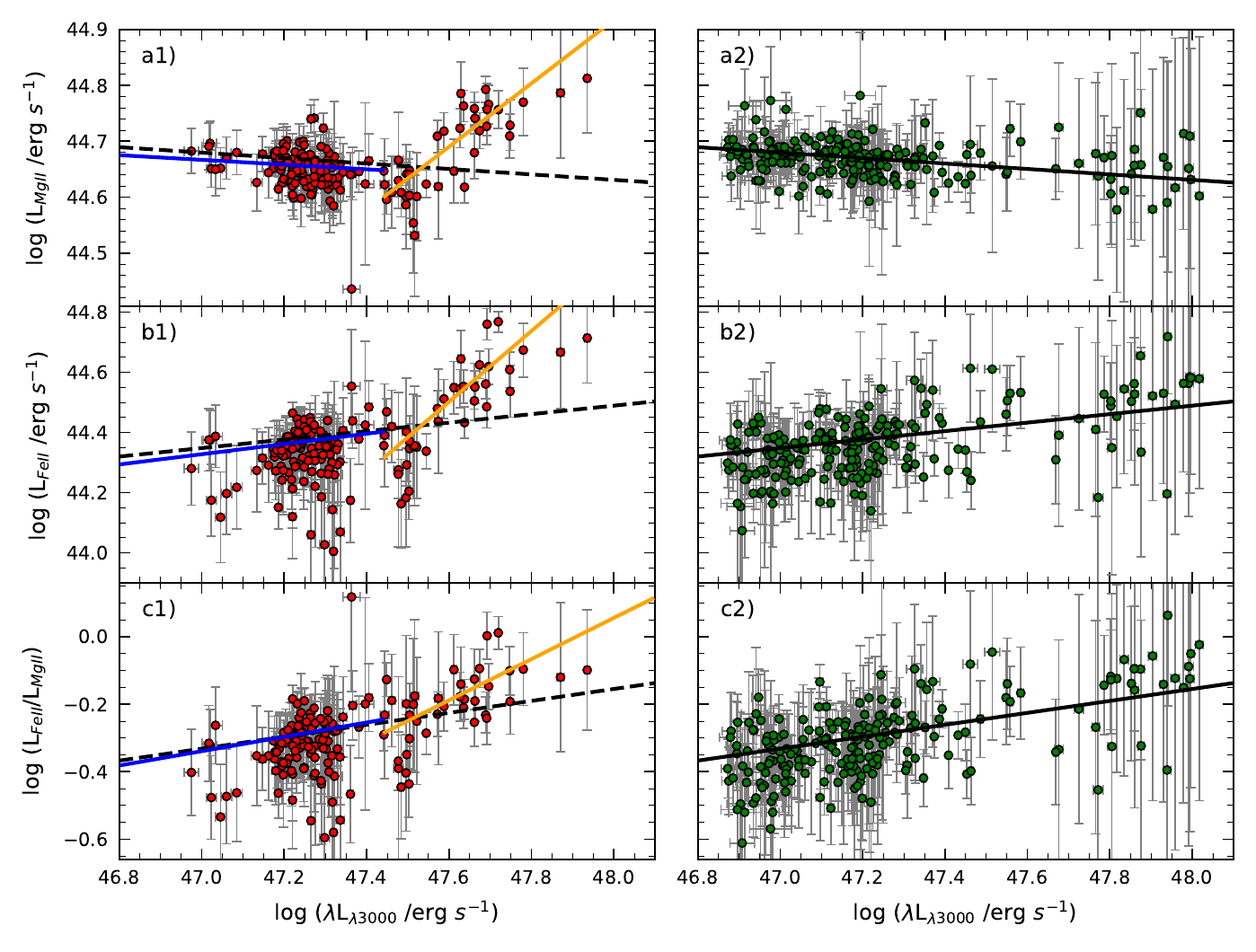}
\caption{Left panel: luminosity relations corresponding to the FP1 described in subsection \ref{Lum_flares}. Right panel: luminosity relations corresponding to the FP2. a1) and a2): variations of the Mg~II $\lambda$2798 \AA\ emission line depending on the $\lambda$3000 \AA\ continuum luminosity. b1) and b2): variations of the UV Fe~II band depending on the $\lambda$3000 \AA\ continuum luminosity. c1) and c2): the ratio 
between the UV Fe~II and the Mg~II $\lambda$2798 \AA\ luminosities varying with the $\lambda$3000 \AA\ continuum luminosity. The linear fit, in a1) and b1), from the beginning to the threshold ($\log (\lambda L_{\lambda 3000}/$erg s$^{-1})=47.44$) is represented by the blue line while the linear fit from the same threshold to the final point is represented by the orange line. In c1), the blue and orange lines represent the quotient of the respective linear fits in a1) and b1). The full linear fit for each luminosity relation of the FP2 is represented by a black solid line in the right panel, and by a black dashed line in the left panel.}
\label{slopes}
\end{center}
\end{figure*}

The analysis done in the last subsection showed that we cannot determine a behavior that describes all the observations. This led us to analyze the luminosity relations separated by the flaring periods.  We selected the luminosity observations during the same periods described in subsection \ref{cc_flares}. These now separated luminosity relations are displayed in Figure \ref{slopes}. The left panel in Figure \ref{slopes} corresponds to the luminosities during the FP1 and the right panel to the FP2. The luminosities of the Mg~II emission line and the UV~Fe~II band do not vary monotonically with the increase of the continuum luminosity during the FP1. However, during the FP2 there is a clear trend. The linear fit for each luminosity relation during the FP2 was no trouble to find, additionally, we calculated their Spearman correlation coefficient. The linear fit of the relation for the Mg~II, displayed in the sub-panel a2) of Figure \ref{slopes}, gives a slope of $-0.05 \pm 0.01$ with a $p_v=2 \times 10^{-5}$ and found a weak anti-correlation ($\rho = -0.27,\ p_v=0.0002$). For the Fe~II, in the sub-panel b2), the slope is $0.15 \pm 0.03$ with a $p_v=0.024$ and we  found a significant correlation ($\rho = 0.52,\ p_v=2\times 10^{-14}$). Finally, for the ratio Fe~II/Mg~II, in the sub-panel c2), the slope found is $0.18 \pm 0.030$ with a $p_v=1\times 10^{-5}$ and the correlation found was significant ($\rho = 0.56,\ p_v=4\times 10^{-17}$) as well. 

From Figure \ref{luminosities}, we can see that at low continuum luminosities until some value (threshold), the observations from the FP1 are indistinguishable from the ones of the FP2, then, they start to act differently at high continuum luminosities. For the luminosity relations of the FP1, in the left panel of Figure \ref{slopes}, we found the linear fit below and above a determined continuum luminosity threshold in the following way: if the dataset has $N$ number of points, we calculated the linear fit for the first 11 points (see subsection \ref{Lum_complete}) and the next $N-11$, then again but for the first 12 and the $N-12$ and so forth, ending with multiple linear fits for below and above the different thresholds. Finally, we found that until $\log (\lambda L_{\lambda 3000}/$erg s$^{-1})=47.44$, for both the Mg~II and Fe~II, the slope is consistent with the one found for their respective datasets of FP2. The black dashed lines on top of the FP1 relations in the left panel of Figure \ref{slopes} correspond to the best fits for the FP2 relations (black solid lines in the right panel). For the Mg~II, displayed in the sub-panel a1) of Figure \ref{slopes}, we found below the threshold a slope of $-0.04 \pm 0.04$ with a $p_v=0.182$ (linear fit not accurate), and above the threshold, a slope of $0.58 \pm 0.07$ with a $p_v=0.724$ (linear fit not accurate). The Spearman correlation test shows a significant anti-correlation ($\rho = -0.32,\ p_v=0.0003$) below the threshold and a strong and significant correlation ($\rho = 0.79,\ p_v=5\times 10^{-9}$) above. For the Fe~II, in the sub-panel b1) of Figure \ref{slopes}, below the threshold the slope found was $0.17 \pm 0.09$ with a $p_v=0.107$ (linear fit not accurate), and above the threshold, we found a slope of $1.18 \pm 0.16$ with $p_v=0.999$ (linear fit not accurate). The Spearman correlation test indicates that below the threshold there is a weak correlation ($\rho = 0.18,\ p_v=0.046$) and above there is a very strong and significant correlation ($\rho = 0.84,\ p_v=4\times 10^{-11}$). The lines depicted for the ratio Fe~II/Mg~II, in sub-panel c1) of Figure \ref{slopes}, are simply the division of the linear fits found for the Mg~II and Fe~II individually, and the Spearman correlation test shows that below the threshold there is a weak but significant correlation ($\rho = 0.28,\ p_v=0.002$) and above there is a strong and significant correlation ($\rho = 0.63,\ p_v=3\times 10^{-5}$).
We sought a luminosity threshold for the FP2, however, the analysis did not show a clear difference between the behaviors at low and high continuum luminosities.

\section{Discussion} \label{sec:disc}

\subsection{Location of the emission region and dominant  emission mechanism of the gamma-rays }

The brightest flares of 2010 and 2014 were both produced by the ejection of a blob (B11 and K14, respectively) from the 43 GHz core. However, the quasi-stationary component C presented a flaring event after the 2014 flare. As showed in Section~\ref{sec:var}, the FP2 was a superposition of multiple events at different locations. Supporting this, the spectral index varied more rapidly during the FP1 than during the FP2 since the latter seems to be more complex, hence, the physical conditions and/or processes that produced the spectra were different at these periods.  This suggests that there are multiple gamma-ray emission regions in 3C~454.3. A change of the location of the gamma-ray emission region of the FSRQ 3C~279 was suggested by \cite{PatinoAlvarez2018} and later observed \citep{PatinoAlvarez2019}, they confirmed the location of a non-stationary gamma-ray emission zone at a distance of at least 42 pc from the core. Furthermore, they suggest that the dominant gamma-ray emission mechanism from this zone is most probably SSC.

Another important aspect to discuss is the dominant gamma-ray emission mechanism during the different periods. \cite{Shah2017} found that to explain the gamma-ray emission of 3C~454.3, it needs to be considered both emission mechanisms, EC and SSC. This has been proposed for FSRQs in general as well \citep{Zacharias2012,ZhangFan2018}. However, for the FSRQ PKS 1510-089, \cite{Chen2012} could not firmly discriminate between neither of the mechanisms for the gamma-ray emission. 
From our cross-correlation analysis, we found for the FP1 time delays of a few days between the gamma-rays, the UV-continuum, and the Fe~II. \cite{Janiak2012} found that delays of a few days (for 3C~279) are well explained when EC is the dominant gamma-ray emission mechanism. If we interpret our delays in the same manner, it would mean that the dominant contribution of seed photons for the FP1 is coming from external sources \citep[the additional BLR, e.g.][]{Leon-Tavares2013}. This is consistent with the results found in the literature \citep{Vittorini2014,Hu2015}.

Recently, \cite{Das2020} using a single-zone model and assuming that the gamma-ray emission region is in the BLR, determined that the dominant mechanism for the FP1 is EC. However, the gamma-ray emission region for this event should be at $\sim$9 pc from the black hole (location of the 43 GHz core). Multizone models have been proposed to address this issue.
One example is when the beamed synchrotron radiation, after scattering off a cloud near the jet trajectory and reentering the jet, can be an important source of photons causing a flaring event \citep{GhiselliniMadau1996,Bottcher1998}. This mirror-like model has been applied to explain the brightest gamma-ray flare of 2010 \citep{Vittorini2014,Tavani2015} with the additional BLR as the mirror and the dominant gamma-ray emission mechanism as EC.

Our cross correlation results for the FP2 show to be consistent with no time delay between the same wavelengths. Thinking along the same lines as for the FP1, the time delay of zero would indicate that the seed photons emission region is co-spatial with the gamma-ray emission region.

For the FP2 we do not see the same significant response from the Mg~II (but for the Fe~II we do) as during the FP1 and there might be multiple gamma-ray emission regions. \cite{Ghisellini2005} applied the spine-sheath model to fit the SEDs of TeV BL Lac objects. This is a multizone alternative to the homogeneous SSC allowing less extreme physical parameters. Each component would see an enhanced radiation field coming from the other component (these components are co-spatial), boosting the IC radiation (with respect to a homogeneous jet). At large distances, the conditions of the jets of FSRQs and BL Lacs could be comparable. This model has been invoked to explain the origin of gamma-ray flares in PKS 1510-089 \citep[e.g.][]{Marscher2010,Park2019}. Particularly, the gamma-ray flare of 2015 of this object was found to have origin at $\sim$10 pc from the base of the jet. \cite{Park2019} attempted to explain this with the spine-sheath model, however, they suggest that for this particular case the jet might consist of multiple layers (not only a fast spine and a slow sheath). A toy model of a spine-sheath structured jet has been developed for FSRQs by \cite{Sikora2016}, they demonstrate that it provides a natural explanation for the hardness and the extreme Compton dominance of the gamma-ray flares, and the weaker fractional flux variations of the optical band than those in gamma-rays. The model allows to follow the dependence of different radiation spectral components on the physical parameters.

Our results indicate that the physical conditions during these two flaring periods were different. The locations and the gamma-ray emission mechanism would have changed for the different flares in the two flaring periods, a similar behavior was found by \cite{PatinoAlvarez2018} for 3C~279.
Finally, to confirm the nature of each significant flare during the FP2, a study of the SED shape using a multizone jet emission model \citep[e.g. multizone turbulent jet,][]{Marscher2014,PeirsonRomani2018,PeirsonRomani2019} with realistic and well defined parameters needs to be performed.

\subsection{The difference in the behavior of Mg~II and Fe~II}

From the cross-correlation analysis, we found a delay of $\sim 640$ days between the $\lambda$3000\AA\ continuum and the Mg~II $\lambda$2798 \AA\ emission line and of $\sim 580$ days for the continuum against the UV Fe~II. \cite{Nalewajko2019} found a similar delay for the Mg~II and used it to calculate the BLR size ($\Delta$r $\sim 0.28$ pc) interpreting the delay as the reverberation of the accretion disk emission. However, we have determined that for most of the UV-continuum light curve the dominant emitting source is the jet. Furthermore, the BLR material that is being ionized does not correspond solely to the canonical BLR, but of an additional BLR surrounding the jet as well \citep[e.g.][]{Leon-Tavares2013}. Hence, we believe these delays should not be used for reverberation mapping since we can not know exactly what is causing the delay.

The luminosity relations showed that the jet is responsible for the major increase in the continuum and, from the light curves we found that the Mg~II $\lambda$2798\AA\ emission line does not increase its flux significantly during the 2014 flare, while the UV Fe~II band does. However, neither the jet or the disk apparently has a major effect in the Mg, and if any, there is an anti-correlation implying that when the continuum increases the Mg decreases. Although, the jet seems to be the source of the continuum that is ionizing the Fe. To explain the anti-correlation of the Mg~II with the continuum, \cite{Nalewajko2019} propose that the continuum has a destructive ionization effect in Mg. But, since we do not see this same effect in the Fe~II, it is important to understand the differences between Mg and Fe to find an explanation. The abundances of Mg and Fe might be unequal, affecting the fluxes of the Mg~II emission line and the UV~Fe~II band, this possibility was studied by \cite{Verner2003}. Another explanation could be the location of Mg and Fe. Since the ionization potentials of the Mg~II and the Fe~II are very similar (7.6 and 7.9 eV, respectively), it is assumed that the Mg~II and the UV Fe~II emission originate in a partially ionized zone of the cloud \citep[][and references therein]{Dietrich2003}. However, \cite{Wills1985} proposed that these are emitted in different zones. The Mg zone being much less extended than the one of Fe. This extended zone of Fe should contain hot regions where the Mg is already ionized emitting Mg~II much less efficiently. \cite{Baldwin2004} propose that the UV Fe~II emission comes from a different component in which Fe is collisionally ionized. This gas would not emit strongly lines of other elements, hence, it would have to be a different BLR component. An additional important difference between Mg and Fe is that they have distinct properties of radiative transfer. The principal difference is the number of resonance lines, Mg~II $\lambda$2798 \AA\ being a doublet while Fe~II involves thousands of transitions. This is due to Fe~II being subject to multiple processes \citep[line fluorescence, Ly~$\alpha$ pumping and turbulence,][]{Verner1999,Verner2003}. As a consequence, \cite{Wills1985} found that Balmer continuum destruction affects the Mg~II much more than the Fe~II. Ergo, for each optically thick Fe~II resonance line there are many escape routes, through conversion to other with smaller optical depths.

\subsection{The difference in the behavior of the spectral features in the different periods}

The cross-correlations of the full dataset had shown an anti-correlation at $\sim 20$ days for the $\lambda$3000\AA\ continuum against the Mg~II $\lambda$2798 \AA\ emission line and the UV Fe~II band. However, when we broke the dataset in the two periods and performed the cross-correlations for each, these anti-correlations disappeared. A new anti-correlation between the UV-continuum and the Mg~II at a delay of $\sim 45$ days appeared, but only for the FP2. We can interpret this again as a destructive ionizing continuum acting upon the Mg.

We found that the response of either of the spectral features changed from the FP1 to the FP2. For the calculation of the linear fits for each luminosity relation of FP1, we determined that the behavior of the spectral features in the FP2 was the same as the behavior during the FP1 but at low continuum luminosities. This result can be interpreted as follows: during the FP2 we see that the Mg~II and Fe~II have a monotonic behavior which tells us that the BLR material is not responding strongly to the increasing continuum. However, we see some increase of the Fe~II. In the FP1 below the threshold we see the same weak response, but, above the threshold we see that both the Mg~II and the Fe~II are responding strongly to the increase of continuum. This indicates that the non-thermal continuum is ionizing efficiently the additional BLR. \cite{Chavushyan2020}, for CTA~102, found a threshold in the luminosity relations between the Mg~II and Fe~II against the continuum. However, the threshold in that source indicates the separation from when the Mg~II and Fe~II are not responding to when they start to present a clear increase in the flux. The authors interpret this behavior as that below the threshold, the material from the canonical BLR is completely ionized (the disk as the dominant source of ionizing continuum), but above the threshold, the continuum (the jet as the dominant source) is being able to ionize the BLR material near the jet. In that case, the ionizing continuum being destructive upon the Mg~II is not observed (the features do not show anti-correlations).

Three possible scenarios might explain why we see a difference in the response of the Mg~II and Fe~II from the FP1 to the FP2. 

The first scenario is that there is a difference in the ionizing spectrum produced by the different physical conditions during the flaring periods. During the FP1 the gamma-ray emission region is at the radio core. Meanwhile, during the FP2 we might have multiple gamma-ray emission locations; ejections from the radio core and a collision at the quasi-stationary component C.

The second scenario is a difference in the state of the additional BLR. During the FP1, we see a strong response of Mg~II and Fe~II to the increase of the ionizing continuum meaning that the BLR material surrounding the jet is present and was able to be ionized \citep{Leon-Tavares2013}. Meanwhile, during the FP2, the behaviors of the luminosity of the Mg~II (decreasing) and the Fe II (increasing) do not display any major changes at high continuum luminosities. In this scenario, the weak responses when the ionizing continuum increases indicate that there is not enough BLR material to be ionized by the jet. The material that was ionized during the FP1 flare could have mostly dissipated due to interaction with the medium by the time of the FP2.

The third scenario is related to the location of the additional BLR. These BLR clouds could be located at a different location during the FP2 making more difficult the ionization of the Mg~II and Fe~II. This additional BLR could have been an important source of seed photons for the IC to produce gamma-rays during the FP1. However, if the BLR clouds by the time of the FP2 are located farther from the radio core, the dominant source of the required seed photons for IC to produce the gamma-rays could be the jet itself (e.g. spine-sheath model).

Finally, it needs to be considered that a combination of these scenarios may as well be a possibility.

\subsection{Implications on the black hole mass calculation}

\begin{figure}[htbp]
\begin{center}
\includegraphics[width=0.48\textwidth]{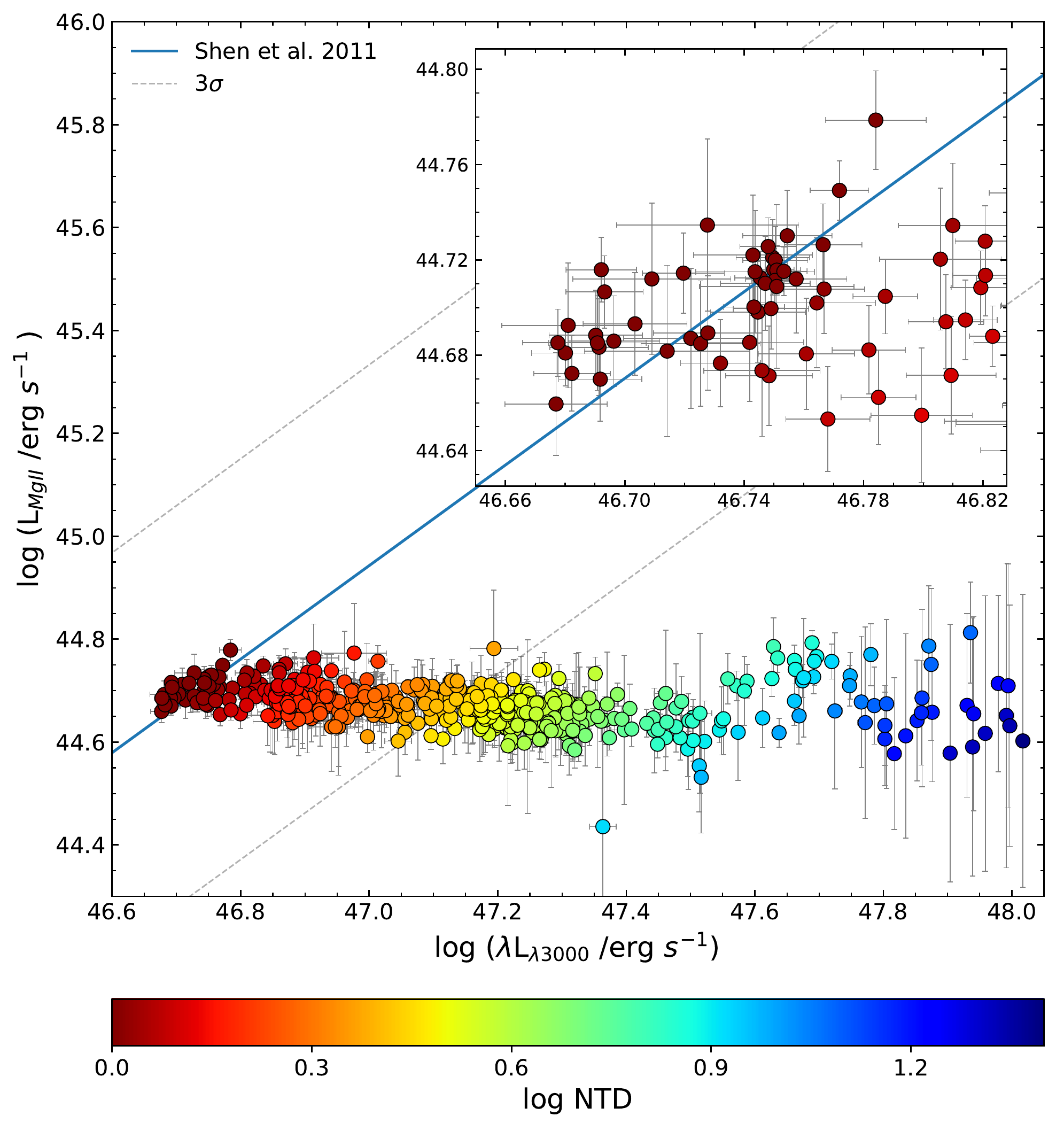}
\caption{Variation of the Mg~II $\lambda$2798\AA\ emission-line luminosity compared to the $\lambda$3000\AA\ continuum luminosity, for 3C~454.3 (our data). The color bar indicates at each observation the NTD value in the logarithmic scale, ranging from 0 to 1.4. Disk dominance ranges between 0 and 0.3 while jet dominance begins after 0.3. The blue solid line and dashed lines represent the bisector fit and its uncertainty at 3$\sigma$, respectively, for a non-blazar sample of \cite{Shen2011}. The embedded figure is a closer look at the points near to the blue line, the NTD values of these points are below 1.3. }
\label{ShenLine}
\end{center}
\end{figure}

In Figure \ref{ShenLine} the relation between the luminosities of the $\lambda$3000 \AA\ continuum and the Mg~II $\lambda$2798 \AA\ emission line of the observations is compared to the relation calculated by \cite{Shen2011} for a non-blazar sample. It is clear that the points do not seem to follow this relation, at least not after NTD~$=2$ (considering the 3$\sigma$ of uncertainty). This indicates that when the continuum is mainly non-thermal, then the relation does not hold. The same behavior was seen for CTA~102 \citep{Chavushyan2020}. This illustrates the important fact that if the optical continuum of a strong radio-loud source is affected by the jet, then, the luminosity of the $\lambda$3000\AA\ continuum will be overestimated. In this case, we are seeing that BLR material close to the base of the jet (beyond the canonical BLR) is involved in the Mg~II emission as well. The reverberation mapping \citep{Blandford_McKee1982,Peterson1993} and single epoch \citep{Vestergaard_Peterson2006} techniques for the calculation of the black hole mass depend crucially of only having a single source of ionization (the accretion disk) and the BLR clouds being in virial equilibrium. Neither of these conditions are fulfilled by 3C~454.3.  We could take an average of the Mg~II luminosity, which do not vary radically with the changes in the UV-continuum, and apply the single epoch technique \citep[e.g.][]{Wang2009}. However, the ideal calculation should be performed by only taking the spectra for which the accretion disk is the dominant source of the continuum (NTD~$<2$), and then proceed with either of the techniques.
 
\section{Summary} \label{sec:sum}

Throughout the years, the blazar 3C~454.3 has been an object of interest due to its frequent periods of high activity. Now, we were able to analyze the spectroscopic data (2009-2018) from the monitoring program of the Ground-based Observational Support of the Fermi Gamma-ray Space Telescope at the University of Arizona. The redshift of this source ($z=0.859$) unlocks the near-UV region of the spectrum allowing us to analyze the Mg~II $\lambda$2798\AA\ emission line and the UV~Fe~II band, as well as, obtain the $\lambda$3000\AA\ continuum. Besides this, we obtained from public sources multiple different wavebands to analyze their respective variability and find the role they play in the flaring events that occurred in 2010 and 2014. The most important results found in this work are listed below:

\begin{itemize}
\item The brightest flares of 2010 and 2014 caused by the ejection of a blob (B11 and K14, respectively) from the 43 GHz core were also observed in the multiple wavelengths discussed in this work. The quasi-stationary component C presented a flaring event after the major flare of 2014 coinciding within the uncertainty with our estimated time of collision between the blob B11 and the quasi-stationary component C. This is an indication that the second flaring period was a superposition of multiple events at different locations.
\item During both flaring periods, the amplitude of the 15 GHz is comparable. Meanwhile, the 1 mm flux had a smaller increase during the second flaring period than during the first. The spectral index, within these frequencies, during the first flaring period had faster changes than during the second. This indicating a more that the physical conditions and/or processes during these periods were different. We suggest that the strange behavior of the spectral index is a result of a superposition of multiple events at different locations.
\item The cross-correlation results show that the emission of the seed photons has a small time lag (of a few days) against the synchrotron emission during the first flaring period. While for the second, the resultant delay is consistent with zero, meaning that the seed photons are coming from the same region as the synchrotron emission itself. To confirm the nature of each individual flare during the second flaring period, a SED study using multizone modeling needs to be performed.
\item The Mg~II and Fe~II behave differently from each other. The Mg~II has an anti-correlation with the continuum while Fe~II correlates positively. Except by the time of the brightest flare of 2010, when both have a strong response at high continuum luminosities. However, Fe~II showed the strongest response. This disparate behavior might be interpreted as BLR clouds composed by different zones or simply because of the different radiative properties of these elements.
\item The Mg~II and Fe~II behave differently during the two flaring periods. In the first flaring period, the emission lines do not behave monotonically to the increase of the continuum. After the found threshold both increase their luminosities correlating with the continuum luminosity. However, during the second flaring period (and at low luminosities of the first flaring period) the Mg~II presents a weak anti-correlation while Fe~II seems to respond positively to the increase of the continuum. This might be explained by the difference in the ionizing spectrum, the state and availability of the BLR, or a different location of the BLR clouds during the second flaring period.
\end{itemize}

The ionizing continuum for almost all the light curve is dominated by the non-thermal emission, and the Mg~II emission is likely to be coming from two different regions, the BLR material inside the inner parsec and the one related to the jet near the 43 GHz core at $\sim$9~pc from the black hole. Therefore, for reverberation mapping and single epoch techniques for the estimation of black hole mass, we strongly recommend using only data for which the accretion disk is the dominant source of continuum (NTD~$<2$), otherwise the errors might be large. For this particular case, the observations with the NTD~$<2$ represent 35\% of all the data.

In the future, we will continue performing multiwavelength studies and analyzing the spectral features of different blazar-type AGNs to further investigate the presence of BLR material beyond the inner parsec of the central engine. This type of studies represent an important tool to find clues about the dominant gamma-ray emission mechanism. It is of particular importance to continue these studies since the determination of the dominant gamma-ray emission mechanism seems to be more complicated than thought before. Which could break the dichotomy of the dominant emission mechanism for FSRQs and BL Lacs.

\acknowledgments 

Acknowledgments. We thank the anonymous referee for the constructive comments that helped to improve the manuscript. This work was supported by CONACyT (Consejo Nacional de Ciencia y Tecnolog\'ia) research grant 280789 (M\'exico). RAA-A acknowledges support from the CONACyT program for PhD studies. This work is supported by the MPIfR-Mexico Max Planck Partner Group led by VMP-A. Data from the Steward Observatory spectropolarimetric monitoring project were used. This program is supported by Fermi Guest Investigator grants NNX08AW56G, NNX09AU10G, NNX12AO93G, and NNX15AU81G \url{http://james.as.arizona.edu/~psmith/Fermi/}. This paper has made use of up-to-date SMARTS optical/near-infrared light curves that are available at \url{www.astro.yale.edu/smarts/glast/home.php}. The 1~mm flux density light curve data from the Submillimeter Array was provided by Mark A. Gurwell. The Submillimeter Array is a joint project between the Smithsonian Astrophysical Observatory and the Academia Sinica Institute of Astronomy and Astrophysics and is funded by the Smithsonian Institution and the Academia Sinica. This research has made use of data from the OVRO 40-m monitoring program \citep{Richards2011} which is supported in part by NASA grants NNX08AW31G, NNX11A043G, and NNX14AQ89G and NSF grants AST-0808050 and AST-1109911. \\

\software{IRAF \citep{Tody1986,Tody1993},
FTOOLS \citep{Blackburn1995},
XSPEC  \citep{Arnaud1996}, 
Fermitools (v 1.0.20), SciPy \citep{2020SciPy-NMeth}
}

\bibliographystyle{aasjournal}  

\bibliography{refs} 

\appendix

\section{Cross-correlation analysis figures}

\figsetstart
\figsetnum{1}
\figsettitle{Cross-correlation function for the complete dataset.}

\figsetend

\begin{figure*}[htbp]
\begin{center}
\includegraphics[width=0.96\textwidth]{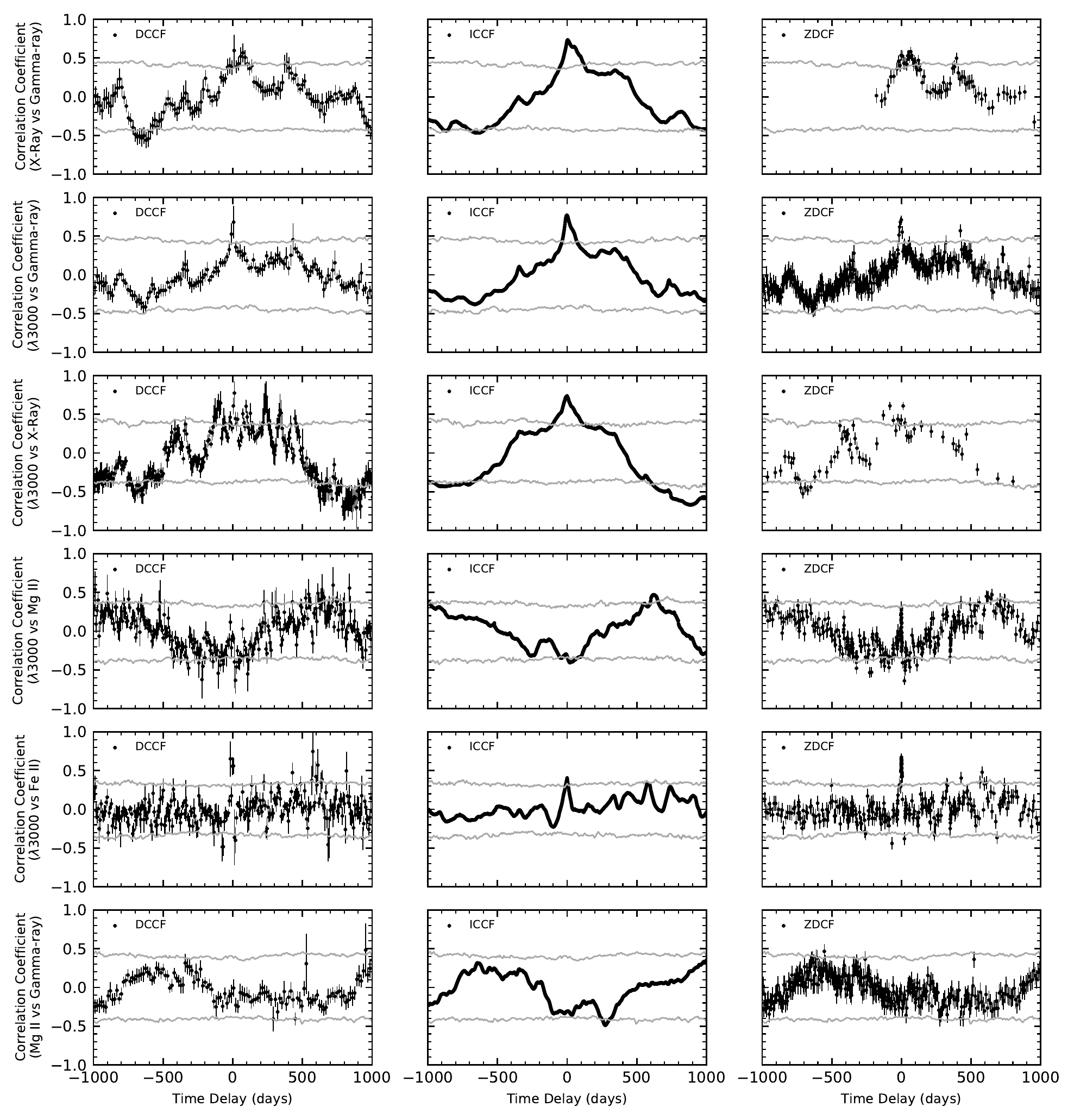}
\caption{Cross-correlations for the complete dataset (see Section \ref{cc_complete}). The complete figure set (6 images) is available in the online journal. Each row corresponds to an analysis done for the two respective features using three different methods. The first method is the Discrete Cross-Correlation Function (DCCF), the second is the Interpolation method (ICCF), and the third method is the Z-transformed Cross-Correlation Function (ZDCF). The significance at the 99\% is represented by a grey plot. The time delays that were found not to be aliases are listed in Table \ref{Table_CC}.}
\label{cc-fig}
\end{center}
\end{figure*}

\figsetstart
\figsetnum{2}
\figsettitle{Cross-correlation function for the FP1.}

\figsetgrpstart
\figsetgrpnum{2.1}
\figsetgrptitle{Cross-correlation function for the FP1 (1)}
\figsetplot{FiguresCC/ccf1_fig1.pdf}
\figsetgrpnote{Cross-correlation analysis for the FP1. Each row corresponds to an analysis done for the two respective features using three different methods. The first method is the Discrete Cross-Correlation Function (DCCF), the second is the Interpolation method (ICCF), and the third method is the Z-transformed Cross-Correlation Function (ZDCF). The significance at the 99\% is represented by a grey plot.}
\figsetgrpend

\figsetgrpstart
\figsetgrpnum{2.2}
\figsetgrptitle{Cross-correlation function for the FP1 (2)}
\figsetplot{FiguresCC/ccf1_fig2.pdf}
\figsetgrpnote{Cross-correlation analysis for the FP1. Each row corresponds to an analysis done for the two respective features using three different methods. The first method is the Discrete Cross-Correlation Function (DCCF), the second is the Interpolation method (ICCF), and the third method is the Z-transformed Cross-Correlation Function (ZDCF). The significance at the 99\% is represented by a grey plot.}
\figsetgrpend

\figsetend

\begin{figure*}[htbp]
\begin{center}
\includegraphics[width=0.96\textwidth]{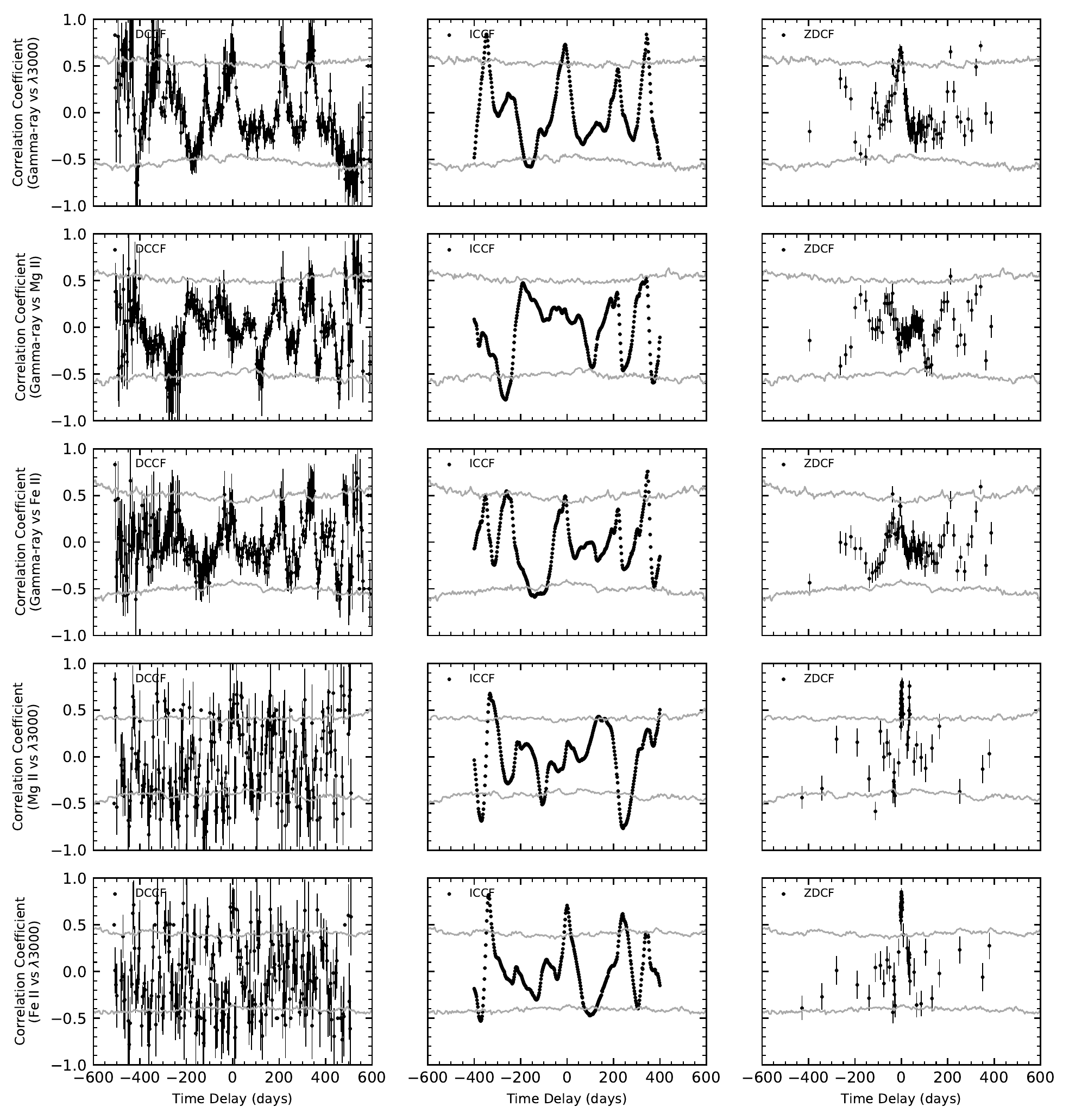}
\caption{Cross-correlations for the FP1 (see Section \ref{cc_flares}). The complete figure set of the FP1 (2 images) is available in the online journal. Each row corresponds to an analysis done for the two respective features using three different methods. The first method is the Discrete Cross-Correlation Function (DCCF), the second is the Interpolation method (ICCF), and the third method is the Z-transformed Cross-Correlation Function (ZDCF). The significance at the 99\% is represented by a grey plot. The time delays that were found not to be aliases are listed in Table \ref{Table_CCflares}.}
\label{cc-figfp1}
\end{center}
\end{figure*}

\figsetstart
\figsetnum{3}
\figsettitle{Cross-correlation function for the FP2.}

\figsetgrpstart
\figsetgrpnum{3.1}
\figsetgrptitle{Cross-correlation function for the FP2 (1)}
\figsetplot{FiguresCC/ccf2_fig1.pdf}
\figsetgrpnote{Cross-correlation analysis for the FP2. Each row corresponds to an analysis done for the two respective features using three different methods. The first method is the Discrete Cross-Correlation Function (DCCF), the second is the Interpolation method (ICCF), and the third method is the Z-transformed Cross-Correlation Function (ZDCF). The significance at the 99\% is represented by a grey plot.}
\figsetgrpend

\figsetgrpstart
\figsetgrpnum{3.2}
\figsetgrptitle{Cross-correlation function for the FP2 (2)}
\figsetplot{FiguresCC/ccf2_fig2.pdf}
\figsetgrpnote{Cross-correlation analysis for the FP2. Each row corresponds to an analysis done for the two respective features using three different methods. The first method is the Discrete Cross-Correlation Function (DCCF), the second is the Interpolation method (ICCF), and the third method is the Z-transformed Cross-Correlation Function (ZDCF). The significance at the 99\% is represented by a grey plot.}
\figsetgrpend

\figsetend

\begin{figure*}[htbp]
\begin{center}
\includegraphics[width=0.96\textwidth]{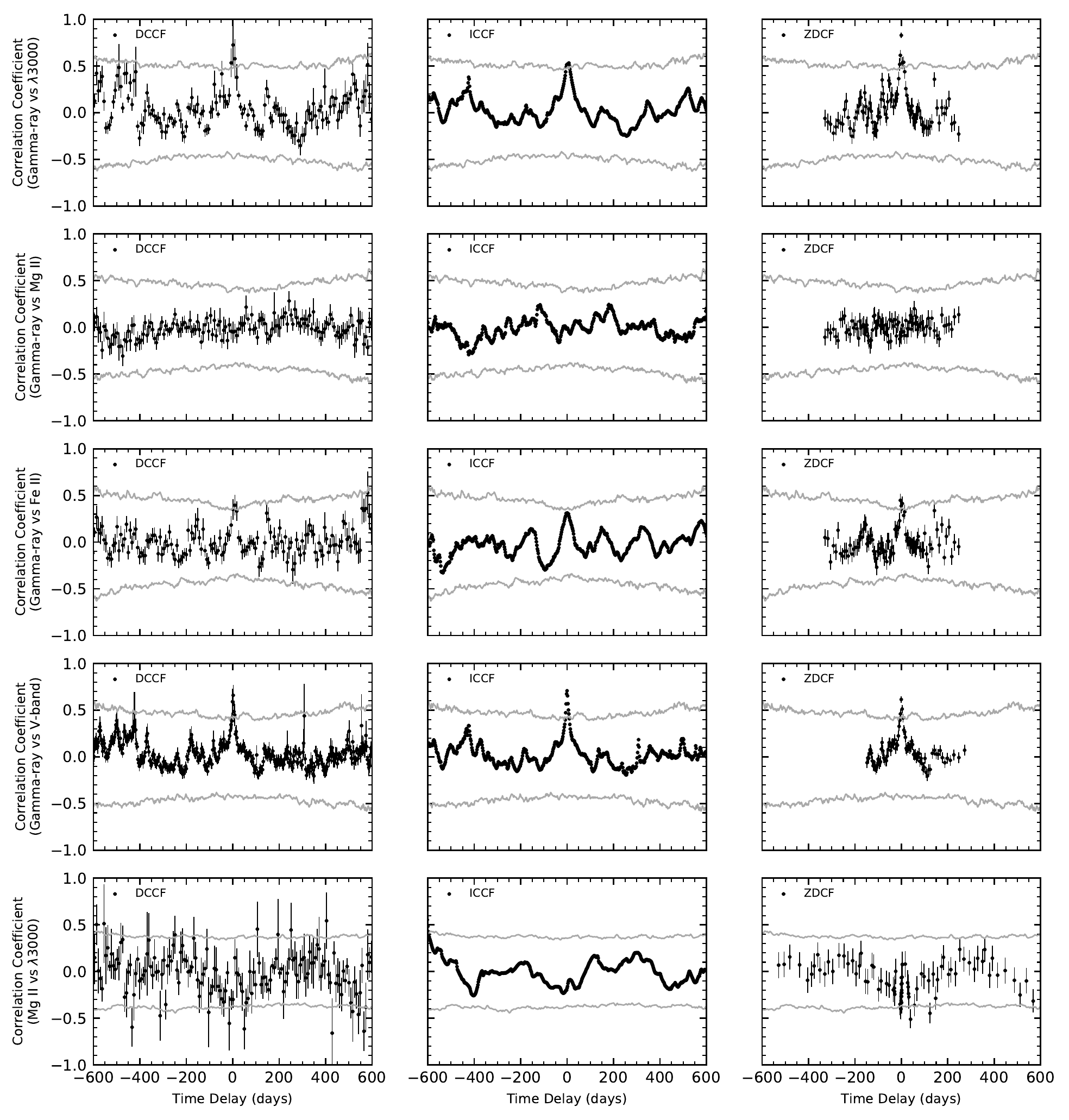}
\caption{Cross-correlations for the FP2 (see Section \ref{cc_flares}). The complete figure set of the FP2 (2 images) is available in the online journal. Each row corresponds to an analysis done for the two respective features using three different methods. The first method is the Discrete Cross-Correlation Function (DCCF), the second is the Interpolation method (ICCF), and the third method is the Z-transformed Cross-Correlation Function (ZDCF). The significance at the 99\% is represented by a grey plot. The time delays that were found not to be aliases are listed in Table \ref{Table_CCflares}.}
\label{cc-figfp2}
\end{center}
\end{figure*}

\end{document}